\documentclass[defaultstyle,10pt,master]{01.thesis}
\usepackage[english]{babel}

\usepackage[latin1]{inputenc}

\usepackage{ulem} % Allows the use of other text emphatizer commands
\usepackage{datetime}
\newdateformat{todaythesis}{%
\monthname[\THEMONTH]  \THEYEAR}

\usepackage{latexsym}

\usepackage{amsmath, amsthm, amssymb, amsfonts, amsbsy}

\usepackage{multirow}
\usepackage{colortbl}
\usepackage{supertabular}
\usepackage{pdflscape}
\usepackage{graphics}
\usepackage{graphicx}
\usepackage{epsfig}
\usepackage[hang,small,bf]{subfigure}
\usepackage{dcolumn}
\usepackage{bm}
\usepackage{booktabs}
\usepackage{rotating}
\usepackage{multirow}

\usepackage[font=small,labelfont=bf,textfont=normalfont]{caption}

\usepackage[square,numbers,sort&compress]{natbib}
\usepackage[printonlyused]{acronym}
\usepackage[titletoc,title,header]{appendix}
\usepackage[noauto]{chappg}

\usepackage{00.extra_functions}

\usepackage[plainpages=false]{hyperref}
\hypersetup{
             colorlinks=false,
             citecolor=red,
             breaklinks=true,
             bookmarksnumbered=true,
             bookmarksopen=true,
             pdftitle={Thesis Title},
             pdfauthor={Author Name},
             pdfsubject={Master Thesis in Biomedical Engineering},
             pdfcreator={Document Creator Name},
             pdfkeywords={Template, Latex, Thesis}}
\usepackage{float}
\usepackage{00.symlist}

\usepackage[top=2.5cm, bottom=2.5cm, inner=2.9cm, outer=2.5cm]{geometry}

\usepackage{fancyhdr}
\fancypagestyle{plain}{% Used in Chapter titles
   \fancyhead{} % get rid of headers
   \fancyfoot[LE,RO]{\bfseries\small\thepage}
}

\fancypagestyle{begin}{%
   \fancyhead{}

   \fancyfoot[LE,RO]{\bfseries\small\thepage}
}
\fancypagestyle{document}{%
	\fancyhf{} 
	\fancyhead[LE]{\bfseries\nouppercase{\leftmark}}
	\fancyhead[RO]{\bfseries\nouppercase{\rightmark}}
	\fancyfoot[LE,RO]{\bfseries\small\thepage}
	\addtolength{\headheight}{2pt} % make space for the rule
}
\fancypagestyle{documentsimple}{%
	\fancyhf{}
	\fancyfoot[LE,RO]{\bfseries\small\thepage}
	\addtolength{\headheight}{2pt} % make space for the rule
}
\graphicspath{{Figures/}}

\def\LL{{\cal L}}
\def\OO{{\cal O}}

\begin{document}
\pagestyle{begin}
\setcounter{page}{1} \pagenumbering{Alph}

% Add PDF bookmark 
\pdfbookmark[0]{Title}{Title}

\thispagestyle{empty}
\begin{flushleft} ~\\ \vspace{-10mm} \hspace{-9mm}  \includegraphics[width=101mm]{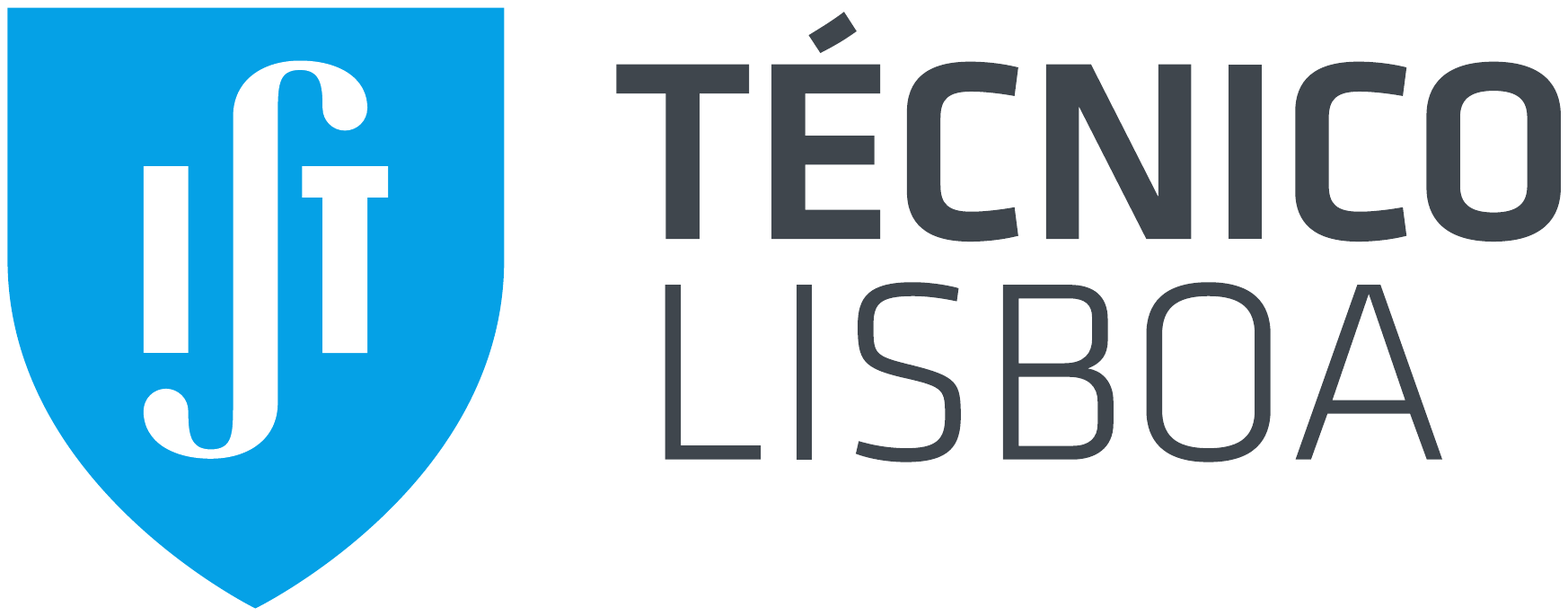} 
\\ \vspace{20mm}
\begin{centering}
\LARGE \textbf{A Nonminimal Coupling Model and its Short-Range Solar System Impact}
\\ \vspace{15mm}
\Large \textbf{Nuno de Albuquerque Emiliano Onofre Castel-Branco}
\\ \vspace{25mm}
\Large Thesis to obtain the Master of Science Degree in
\\ \vspace{2mm}
\LARGE \textbf{Engineering Physics}
\\ \vspace{10mm}

\begin{tabular}{lll}
\large Supervisors:    & \large Prof. Doutor V\'itor Manuel dos Santos Cardoso  & \large \\
\large & \large Prof. Doutor Jorge Tiago Almeida P\'aramos & \large \\  
\end{tabular}

\vspace{10mm}

\Large \textbf{Examination Committee} \large

\vspace{5mm}

\begin{tabular}{lll}
 Chairperson:		& Prof. Doutora Ana Maria Vergueiro Monteiro Cidade Mour\~ao &  \\ 
\large Supervisor: 		& Prof. Doutor Jorge Tiago Almeida P\'aramos & \large \\  
\large Members of the Committee:	 				& Prof. Doutor Amaro Jos\'e Rica da Silva & \large \\
\large 	 				& Prof. Doutor Francisco Sab\'elio Nobrega Lobo   & \large \\
\end{tabular}
 
\vspace{10mm}

\Large \textbf{April 2014} \\
\end{centering}
\let\thepage\relax
\end{flushleft}
\pagebreak

\clearpage
% Since I am using double sided pages, the second page should be white.
% Remember that when delivering the dissertation, IST requires for the cover to appear twice.

\thispagestyle{empty}
\cleardoublepage

\setcounter{page}{1} \pagenumbering{roman}

\baselineskip 18pt % line spacing: -12pt for single spacing
                   %               -18pt for 1 1/2 spacing
                   %               -24pt for double spacingnts} % Uncomment for normal cover with IST logo.
\thispagestyle{empty}
%\hbox{} \vfill

\vspace*{\fill}
\begingroup
\centering
\begin{flushleft}
\small ``I have never found a better

expression than `religious'

for this trust in the rational

nature of reality and of its peculiar

accessibility to the human mind.

\quad

Where this trust is lacking,

science degenerates into

an uninspired procedure.

\quad

Let the devil care if

the priests make capital out of this.

There is no remedy for that.''

\vspace{10mm}  

 - Albert Einstein (1879-1955)

\quad

\textit{Lettres a Maurice Solovine reproduits en facsimile et traduits en fran\c{c}ais}

Paris: Gauthier-Vilars, 1956, pp. 102-103.

\end{flushleft}
\endgroup
\vspace*{\fill}

\clearpage
\thispagestyle{empty}
\cleardoublepage

%\pdfbookmark{Acknowledgments}{Acknowledgments}
%\addcontentsline{toc}{chapter}{Acknowledgments}
\begin{acknowledgments} 

This Master thesis is for me the conclusion of the Physics graduation I began in the \textit{Instituto Superior T\'ecnico} (IST) after I left High School. For this conclusion I am grateful, for it is a door that closes in my life and a bigger one that opens.

These years have been full of great things which could not have happened if it were not for many good friends, colleagues and professors alike. Of all these, I would like to thank at first my supervisor Professor Jorge P\'aramos, who has been accompanying me for at least two years and has taught me that great competence can be achieved at a very young age. Word is also due for Professor V\'itor Cardoso with whom I learned much on the Astrophysics course he lectured me and who accepted this challenge of being my supervisor and for Professor Orfeu Bertolami who was a great help with many fruitful discussions in the development of the last and major part of this work.

This thesis began with a scholarship granted to me by the \textit{Laboratorio Nazionale di Frascati}, Rome, from the \textit{Istituto Nazionale di Fisica Nucleare}, where I researched from November 2012 to April 2013. Simone Dell'Agnello, leader of the group where I was inserted, always encouraged me to do the best work possible and to aim high when doing Science and for this I am thankful. Also, most of the things I learned related to Physics in my Roman months were from Riccardo March, a great and experienced mathematician, who always had time to explain me even the most trivial details of our work.

I would also like to acknowledge all the professors of the Department of Physics of the IST who lectured me and also all the friends I have made among my colleagues. During these years many friends outside the Physics graduation were essential too, in making me grow at a personal and an intellectual level. As they are so many I will only mention Fr. Hugo Santos, chaplain of the Catholic University of Portugal, who besides introducing me to many great friendships, led me to the long and amazing path of searching for the Truth, of which Science is an essential part. A highlight should be given to Leonor's work in helping me review the English correctness of the text.

Finally I would like to thank my amazing parents without whom nothing of this would be possible. I would like to thank my Mother for her understanding and support at all levels and my Father whose scientific and other personal interests were essential for me to choose Physics as a university program. I would also like to thank my grandparents for all the talks and time we spent together during the making of the thesis and, last but not least, my Brother and Sister who have a gift to make me laugh and help me remember that family is the central part of anyone's life.

\end{acknowledgments}

\cleardoublepage

\begin{resumo}
%\addcontentsline{toc}{chapter}{Resumo}

O objectivo deste trabalho \'e apresentar os efeitos de um modelo de acoplamento n\~ao-m\'inimo da gravidade no Sistema Solar, num regime de curto alcance. Por esta razão, este estudo só é válido quando a contribuição cosmológica é considerada irrelevante. A ac\c{c}\~ao do modelo inclui duas fun\c{c}\~oes $f^1(R)$ e $f^2(R)$ da curvatura escalar de Ricci $R$, onde a segunda multiplica o Lagrangeano da mat\'eria.

Atrav\'es de uma expans\~ao de Taylor em torno de $R=0$ para ambas as fun\c{c}\~oes $f^1(R)$ e $f^2(R)$, descobriu-se que a m\'etrica \`a volta de um objecto esf\'erico \'e uma perturba\c{c}\~ao da aproxima\c{c}\~ao de campo fraco da m\'etrica de Schwarzschild. A componente $tt$ da métrica, um termo Newtoniano com um termo perturbativo de Yukawa, \'e condicionada atrav\'es dos resultados observacionais dispon\'iveis.

Em primeiro lugar, verifica-se que este efeito é anulado quando as escalas de massa caracter\'isticas de cada fun\c{c}\~ao $f^1(R)$ e $f^2(R)$ s\~ao id\^enticas. Para além disto, a conclus\~ao \'e que o acoplamento n\~ao-m\'inimo s\'o afecta a for\c{c}a da contribui\c{c}\~ao de Yukawa e n\~ao o seu alcance e que o modelo de Starobinsky para a infla\c{c}\~ao n\~ao está limitado experimentalmente. Mais ainda, o efeito da precess\~ao geod\'etica, obtida tamb\'em da perturba\c{c}\~ao radial da m\'etrica, revela n\~ao ter relev\^ancia para os limites obtidos.

O trabalho original apresentado nesta tese segue ao pormenor a Ref. \cite{nunocb}.

\end{resumo}

\begin{palavraschave}
\noindent Teorias $f(R)$; Acoplamento n\~ao m\'inimo; Modifica\c{c}\~oes Yukawa \`a lei do quadrado inverso; Sistema Solar
\end{palavraschave}
\clearpage
\thispagestyle{empty}
\cleardoublepage
%\pdfbookmark{Abstract}{Abstract}
%\addcontentsline{toc}{chapter}{Abstract}

\begin{abstract}

The objective of this work is to present the effects of a nonminimally coupled model of gravity on a Solar System short range regime. For this reason, this study is only valid when the cosmological contribution is considered irrelevant. The action functional of the model involves two functions $f^1(R)$ and $f^2(R)$ of the Ricci scalar curvature $R$, where the last one multiplies the matter Lagrangian.

Using a Taylor expansion around $R=0$ for both functions $f^1(R)$ and $f^2(R)$, it was found that the metric around a spherical object is a perturbation of the weak-field Schwarzschild metric. The $tt$ component of the metric, a Newtonian plus a Yukawa perturbation term, is constrained using the available observational results.

First it is shown that this effect is null when the characteristic mass scales of each function $f^1(R)$ and $f^2(R)$ are identical. Besides, the conclusion is that the nonminimal coupling only affects the Yukawa contribution strength and not its range and that the Starobinsky model for inflation is not experimentally constrained. Moreover, the geodetic precession effect, obtained also from the radial perturbation of the metric, reveals to be of no relevance for the constraints.

The original work presented in this thesis closely follows Ref. \cite{nunocb}.

\end{abstract}
\clearpage
\thispagestyle{empty}
\cleardoublepage
% This is required for the fancy chapters
\dominitoc
\dominilof
\dominilot

%%%%%%%%%%%%%%%%%%%%%%%%%%%%%%%%%%%%%%%%%%%%%%%%%%%%%%%%%%%%%%%%%%%%%%
% List of contents
%\renewcommand{\baselinestretch}{1}
\pdfbookmark[0]{Index}{index}
\pdfbookmark[1]{Contents}{toc}
\tableofcontents
% \contentsline{chapter}{References}{\pageref{bib}}
\clearpage
\thispagestyle{empty}
\cleardoublepage
%\renewcommand{\baselinestretch}{1.5}
%%%%%%%%%%%%%%%%%%%%%%%%%%%%%%%%%%%%%%%%%%%%%%%%%%%%%%%%%%%%%%%%%%%%%%
% List of figures
%\addcontentsline{toc}{chapter}{List of Figures}
\pdfbookmark[1]{List of Figures}{lof}
\listoffigures
\clearpage
\thispagestyle{empty}
\cleardoublepage

%%%%%%%%%%%%%%%%%%%%%%%%%%%%%%%%%%%%%%%%%%%%%%%%%%%%%%%%%%%%%%%%%%%%%%
% List of tables
%\pdfbookmark[1]{List of Tables}{lot}
%\listoftables
%\clearpage
%\thispagestyle{empty}
%\cleardoublepage

% %%%%%%%%%%%%%%%%%%%%%%%%%%%%%%%%%%%%%%%%%%%%%%%%%%%%%%%%%%%%%%%%%%%%%%
% % List of algorithms
% Requires packages algorithmic, algorithm
% \pdfbookmark[1]{List of Algorithms}{loa}
% \listofalgorithms
% \cleardoublepage
\acresetall
% %%%%%%%%%%%%%%%%%%%%%%%%%%%%%%%%%%%%%%%%%%%%%%%%%%%%%%%%%%%%%%%%%%%%%%
 % List of acronyms
\pdfbookmark[1]{List of Acronyms}{loac}
%\addcontentsline{toc}{chapter}{Abbreviations}

\chapter*{Abbreviations}

% See more at http://staff.science.uva.nl/~polko/HOWTO/LATEX/acronym.html

\begin{acronym}
\acro{GR}{General Relativity}
\end{acronym}

\begin{acronym}
\acro{EFE}{Einstein Field Equations}
\end{acronym}

\begin{acronym}
\acro{FLRW}{Friedmann-Lema\^itre-Robertson-Walker}
\end{acronym}

\begin{acronym}
\acro{EOS}{equation of state}
\end{acronym}

\begin{acronym}
\acro{BD}{Brans-Dicke}
\end{acronym}

\begin{acronym}
\acro{SRA}{Slow-Roll Approximation}
\end{acronym}

\begin{acronym}
\acro{NMC}{nonminimal coupling}
\end{acronym}

\begin{acronym}
\acro{WMAP}{Wilkinson Microwave Anisotropy Probe}
\end{acronym}

\begin{acronym}
\acro{CMB}{Cosmic Microwave Background}
\end{acronym}

\begin{acronym}
\acro{ISL}{Inverse-Square Law}
\end{acronym}

\begin{acronym}
\acro{PPN}{Parametrized Post-Newtonian}
\end{acronym}

\begin{acronym}
\acro{GPB}{Gravity Probe B}
\end{acronym}

\begin{acronym}
\acro{LLR}{Lunar Laser Ranging}
\end{acronym}

\clearpage
\thispagestyle{empty}
\cleardoublepage

%%%%%%%%%%%%%%%%%%%%%%%%%%%%%%%%%%%%%%%%%%%%%%%%%%%%%%%%%%%%%%%%%%%%%%
% List of symbols
%\pdfbookmark[1]{List of Symbols}{los}

%\listofsymbols

\clearpage
\thispagestyle{empty}

\cleardoublepage
% Pages number is starting now with arabic style... until now it was on roman mode
\pagenumbering{arabic} \setcounter{page}{1}
\baselineskip 18pt
\pagestyle{documentsimple}%Simple head
% %%%%%%%%%%%%%%%%%%%%%%%%%%%%%%%%%%%%%%%%%%%%%%%%%%%%%%%%%%%%%%%%%%%%%%
% The Introduction:
% %%%%%%%%%%%%%%%%%%%%%%%%%%%%%%%%%%%%%%%%%%%%%%%%%%%%%%%%%%%%%%%%%%%%%%
\chapter{Introduction}
\label{cap:int}

\section{General Relativity}
\label{sec:GR}

To study gravity, physicists rely on the theory of \ac{GR}, as they have done for almost a hundred years. \ac{GR} was developed by Einstein in 1916 and since then it has been able to demonstrate itself as a solid theory of physics. After its theoretical framework was developed, the observational confirmations didn't take long to appear: several experiments have been done to confirm it \cite{will}, such as the measure of the perihelion of Mercury, the detection of the bending of light or the E\"otv\"os torsion balances experiment, just to name a few of the classics. Besides these, also today there are people working on very recent General Relativity tests and its most dramatic consequences, such as the LAGEOS \cite{lageos} and Lunar Laser Ranging \cite{LLR} collaborations.

One of the main tenets of \ac{GR} is the relation between matter and geometry, where gravity appears not as a force field but as a deformation of the spacetime. This is mathematically represented in the \ac{EFE}, which, as all field equations in physics, are obtained from an action principle. Indeed, consider the Einstein-Hilbert action

\begin{equation}
S = \int \left[\kappa \left(R - 2\Lambda \right) + \LL_m \right] \sqrt{-g} \text{ d}^4 x,
\label{eq:einstein-hilbert_action}
\end{equation}

\noindent where $R$ is the Ricci scalar, $\LL_m$ is the matter lagrangian, $\Lambda$ stands for the Cosmological Constant, $g$ is the determinant of the metric and $\kappa$ is defined as $\kappa = c^4 / (16\pi G)$, with $G$ as Newton's constant of gravity  and $c$ as the velocity of light in a vacuum.

Hence, by the principle of least action, the variation of the action $\delta S = 0$ with respect to the metric $g^{\mu\nu}$ yields the \ac{EFE},

\begin{equation}
G_{\mu\nu} + \Lambda g_{\mu\nu} = \dfrac{1}{2\kappa} T_{\mu\nu},
\label{eq:einstein_field_equations}
\end{equation}

\noindent where $G_{\mu\nu} \equiv R_{\mu\nu} - \dfrac{1}{2} g_{\mu\nu}R $ is the Einstein tensor and $T_{\mu\nu}$ is the energy-momentum tensor of matter, clearly showing the already stated relation between matter and geometry. Applying the Bianchi identities, it is found that these equations satisfy the conservation of energy,

\begin{equation}
\nabla^\mu G_{\mu\nu} = 0 \Rightarrow \nabla^\mu T_{\mu\nu} = 0.
\label{eq:conversation_energy_GR}
\end{equation}

\section{Standard Cosmology and Inflation}
\label{sec:standard_cosmology}

Following the Cosmological Principle, which states that the universe is homogeneous and isotropic, the universe can be said to have a \ac{FLRW} metric \cite{walker,robertson}:

\begin{equation}
ds^2 = - c^2 dt^2 + a^2(t) \left( \dfrac{dr^2}{1 - K r^2} + r^2 d\Omega^2 \right)
\label{eq:metric_flrw},
\end{equation}

\noindent where $a(t)$ is the cosmological scale factor, $d\Omega^2$ is the line element for the 2-sphere and $K$ is related to the curvature of the spatial section of spacetime, where the spatial coordinates are comoving coordinates.

One may consider that at very large scales the universe can be described as being a perfect fluid, endowed with an energy-momentum tensor

\begin{equation}
T_{\mu\nu} = \left(\rho + \dfrac{p}{c^2} \right) u_\mu u_\nu + p g_{\mu\nu},
\label{eq:tensor_perfect_fluid}
\end{equation}

\noindent where $\rho$ is the energy density, $p$ is the pressure and $u_\mu$ is the 4-velocity vector. Inserting this tensor and the metric \eqref{eq:metric_flrw} into the \ac{EFE} \eqref{eq:einstein_field_equations} gives the cosmology structure equations:

\begin{eqnarray}
\dot{a}^2 + K c^2 &=& \dfrac{a^2}{3} \left( 8 \pi G \rho + \Lambda c^2 \right), \label{eq:friedmann_eq_1}\\
\dfrac{\ddot{a}}{a} &=& - \dfrac{4 \pi G}{3}\left( \rho + \dfrac{3 p}{c^2}\right) + \dfrac{\Lambda c^2}{3},
\label{eq:cosmology_structure_equations}
\end{eqnarray}

\noindent where $\dot{a}=da/dt$. The first of these equations is called the Friedmann equation. Through a simple manipulation, both these equations can be written in the form of the conservation of energy:

\begin{equation}
\dot{\rho} + 3 H \left( \rho c^2 + p \right) \Leftrightarrow
\dfrac{d}{dt} \left( \rho c^2 a^3 \right) + p \dfrac{d}{dt} a^3 = 0,
\label{eq:conservation_energy_cosmology}
\end{equation}

\noindent which can also be directly obtained from Eq. \eqref{eq:conversation_energy_GR}. The Hubble parameter is extremely useful to characterize the evolution of the universe and is defined as

\begin{equation}
H = \dot{a} / a.
\label{eq:hubble_parameter}
\end{equation} 

The above equations \eqref{eq:friedmann_eq_1} and \eqref{eq:cosmology_structure_equations} yield a full description of the structure and development of the universe if an \ac{EOS} is known. The simplest approach is to consider a barotrope, described by the equation

\begin{equation}
p = w \rho,
\label{eq:eos_standard}
\end{equation}

\noindent where $w$ is a constant independent of time. Notice that inserting this \ac{EOS} into the conservation of energy \eqref{eq:conservation_energy_cosmology} yields

\begin{equation}
\dfrac{\dot{\rho}}{\rho} = - 3 \left( 1 + w \right) H.
\end{equation}

\noindent This expression can be integrated to obtain the solution of $\rho$,

\begin{equation}
\rho (a) = \rho_0 a^{-3 \left(1 + w \right)},
\end{equation}

\noindent where $\rho_0$ is the value of the energy density at the present time $t_0$, when $a(t_0) \equiv a_0$.

The universe underwent different stages of evolution, characterized by the dominance of different kinds of matter, in the so-called radiation, matter and dark energy eras. In the radiation era the Universe was dominated by electromagnetic radiation or massive particles travelling at velocities close to the speed of light, so that its \ac{EOS} is

\begin{equation}
p_r = \dfrac{1}{3} \rho_r.
\end{equation}

\noindent The energy density in the radiation era falls off as $\rho_r \propto a^{-4}$. As for the matter era, dominated by nonrelativistic particles with approximately no collisions, matter has essentially zero pressure $p_m =0$, so that $w_m=0$. In this case the energy density falls off as $\rho_m \propto a^{-3}$.

Knowing how the energy density behaves with the scale factor, it is possible to derive the function $a(t)$ through the Friedmann equation \eqref{eq:friedmann_eq_1} and the Hubble parameter definition \eqref{eq:hubble_parameter}. Considering a flat model $K=0$ with $\Lambda = 0$, in the radiation era the scale factor becomes $a(t) \propto t^{1/2}$ and in the matter era, $a(t) \propto t^{2/3}$. It should also be mentioned that when a non-vanishing cosmological constant $\Lambda \neq 0$ dominates, so that $\rho_\Lambda = -p_\Lambda$ (i.e. $\omega_\Lambda =-1$), one obtains
$\rho_\Lambda \propto a$, so that $a(t) \propto \exp (Ht)$. When the universe behaves according to the latter model it means that it is going through a De Sitter phase, which characterizes the dark energy era.

The standard Big-Bang model, described by Eqs. \eqref{eq:cosmology_structure_equations} and \eqref{eq:friedmann_eq_1}, has seen many of its predictions confirmed by cosmological observations, such as the galaxies red-shift or the detection of the \ac{CMB} \cite{weinberg}. Nonetheless, some observations showed that there were three problems in this model. They were the horizon problem, related to the explanation of how regions of the universe not causally related were in thermal equilibrium, as shown by the \ac{CMB} Radiation, the flatness problem, from the extreme fine-tuning necessary to explain how the universe is as spatially flat as it looks, and the monopole problem, indicating that these exotic entities should dominate the universe in the present time, contrary to observations. These three problems were solved in a single theory, called inflation \cite{guth}.

The theory of inflation comes directly from the solution to the flatness problem. Defining the critical density as 

\begin{equation}
\rho_c(t) = 3 H^2 \slash 8 \pi G,
\label{eq:critical_density}
\end{equation}

\noindent and neglecting the cosmological constant term, the Friedmann equation \eqref{eq:friedmann_eq_1} can be written as

\begin{equation}
\left| \Omega - 1 \right| = \dfrac{|K|c^2}{a^2 H^2} = \dfrac{\left| K \right| c^2}{\dot{a}^2} ,
\label{eq:friedmann_eq_2}
\end{equation}

\noindent where $\Omega$ is the density parameter defined as $\Omega = \rho \slash \rho_c$. The standard cosmological model predicts that the $r.h.s.$ of the above equation should increase with time, hence $\Omega$ should become larger with time since the Big Bang, but observations yield an $\Omega \sim 1$, i.e. the universe is flat ($K=0$). The only way to counter this would be for the $r.h.s.$ to decrease with time:

\begin{equation}
\dfrac{\left| K \right| c^2}{\dot{a}^2} \text{ should decrease } \Rightarrow \dot{a} \text{ should increase } \Rightarrow \ddot{a} > 0,
\label{eq:inflation_condition}
\end{equation}

\noindent which is the inflation condition.

The origin of inflation is explained by means of a scalar field named inflaton, $\phi$, whose effective energy density and pressure are written as

\begin{equation}
\rho_\phi = \dfrac{\dot{\phi}^2}{2} + V(\phi), \qquad\qquad\qquad
p_\phi = \dfrac{\dot{\phi}^2}{2} - V(\phi),
\label{eq:energy_pressure_phi}
\end{equation}

\noindent where the potential $V(\phi)$ depends on the inflation model in question. Moreover, the inflation condition means that the potential energy must dominate in order to obtain a phase of accelerated expansion,

\begin{equation}
\ddot{a} > 0 \Leftrightarrow p < -\rho c^2 \slash 3 \Leftrightarrow \dot{\phi}^2 < V(\phi),
\label{eq:potential_dominates_inflation}
\end{equation}

\noindent where the second cosmology structure equation \eqref{eq:cosmology_structure_equations} was used.
This is the so-called \ac{SRA}, which states that the kinetic term can be neglected with respect to the potential term. Notice that if this condition applies, then by Eq. \eqref{eq:energy_pressure_phi}, the \ac{EOS} becomes similar to that of a Cosmological Constant, $\rho_\phi \simeq -p_\phi$. Inserting the inflaton energy density and pressure \eqref{eq:energy_pressure_phi} into the Friedmann equation \eqref{eq:friedmann_eq_1} and the energy conservation \eqref{eq:conservation_energy_cosmology} gives the inflation structure equations:

\begin{eqnarray}
H^2 &=& \dfrac{8\pi G}{3} \left[ V(\phi) + \dfrac{\dot{\phi}^2}{2}\right], \\
\ddot{\phi} +3 H \dot{\phi} &=& - V'(\phi), \label{eq:conservation_eq_inflation}
\end{eqnarray}

\noindent where $V'(\phi) = dV \slash d\phi$. By the \ac{SRA}, these equations become

\begin{eqnarray}
H^2 &\simeq & \dfrac{8 \pi G}{3} V(\phi), \\
3 H \dot{\phi} &\simeq & - V'(\phi).
\end{eqnarray}

\noindent It is also possible to define the slow-roll parameters as

\begin{eqnarray}
\epsilon (\phi) &=& \dfrac{1}{16 \pi G} \left( \dfrac{V'}{V} \right)^2 \ll 1, \\
\left|\eta (\phi) \right| &=& \left| \dfrac{1}{8\pi G} \dfrac{V''}{V} \right| \ll 1,
\end{eqnarray}

\noindent which measure the slope and the curvature of $V(\phi)$, respectively. Their smallness is necessary to satisfy the \ac{SRA}.

The number of $e$-folds between an initial $\phi_i$ and final state $\phi_f$ of the inflation is determined by

\begin{equation}
N = \ln \dfrac{a(t_f)}{a(t_i)} = \int^{t_f}_{t_i} H dt = -8 \pi G \int^{\phi_f}_{\phi_i} \dfrac{V}{V'} d\phi,
\end{equation}

\noindent where $dt=d\phi/\dot{\phi}$ was used. It is known that for inflation to overcome the three original problems of the standard cosmological model, $N$ should be no less than $70$ \cite{felice}.

\begin{figure}[ht]
\centering
\includegraphics[width=0.7\textwidth]{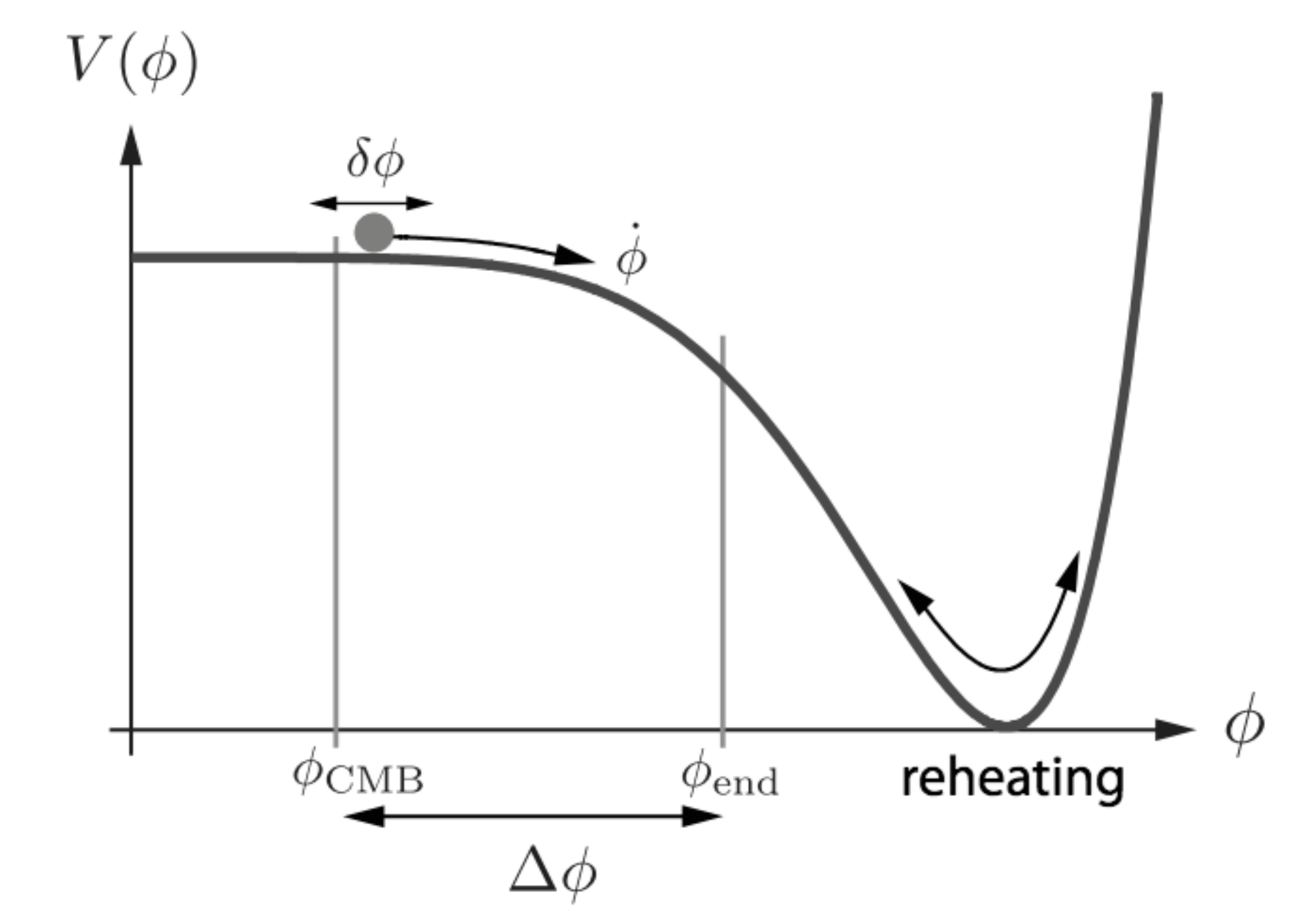}
\caption{Typical inflation potential. Adapted from Ref. \cite{reheatingpic}.}
\label{fig:inflation_potential}
\end{figure}

The duration of inflation is the time when the \ac{SRA} is valid, a period in which the energy density and temperature of the universe are extremely low. As such, at the end of inflation, the potential $V(\phi)$ loses its strength and the universe must go through a reheating process, as the exponential expansion has left the Universe at essentially zero temperature. This is achieved through an oscillation of the inflaton in a potential well, as shown in figure \ref{fig:inflation_potential}.

Quantum field theory explains that an oscillating electric field produces electron-positron pairs; similarly, the oscillation of the field $\phi$ produces its own particles, with the energy that it is losing until the end of the oscillations. This will happen until the temperature needed to go back to the standard cosmological model is reached. Such a process is explained in a more analytical way through the conservation of energy inflation equation \eqref{eq:conservation_eq_inflation}, by adding the term $\Gamma_\phi$, which is the decay range of $\phi$:

\begin{equation}
\ddot{\phi} + 3H\dot{\phi} + \Gamma_\phi \dot{\phi} + V'(\phi) = 0.
\label{eq:oscillation_reheating}
\end{equation}

\noindent This can be interpreted as a friction term, where the energy lost by the inflaton is transferred to the particles produced, thus reheating the Universe.

Without entering into much details, inflation has already some observational evidences, such as the last year results from the Planck experiment \cite{planckinflation} and the very recent clues of evidence of gravitational waves \cite{bicep2} (although some doubts have arisen as to the validity of these findings \cite{bicep2not1,bicep2not2}).

\section{Modern Problems}
\label{sec:mod_problems}

Besides the exact origins of inflation, other problems still remain to be explained, such as the relation between quantum mechanics and gravity (check Ref. \cite{quantumgravity} for a review on the subject) and the nature of dark matter and dark energy. These concepts arise as an explanation to an observational effect that is contrary to the theoretical prevision of \ac{GR}. Dark Matter is inferred, for example, from the anomalous flattening of the galaxy rotation curves: at the periphery of galaxies, the rotation velocity of each galaxy should be much slower than observed, hence there should be some dark ($i.e.$ unseen, electromagnetically invisible) matter that yields this gravitational effect. There are other observational evidences that show this effect, which can be found in Ref. \cite{darkmatter1}, where some of the candidates to dark matter and detection methods are also explained. 

Dark energy, on the other hand, is related to the cosmological scale of the universe. In the late 1990s, the observation of the red-shift of distant supernovae became the main evidence for the accelerated expansion of the universe \cite{perlmutter,riess}. Unexplained by visible matter in the standard cosmological model, many solutions have been proposed, commonly known as dark energy. Indeed, as the moving force of the accelerated expansion, recent observations have concluded dark energy to be $68,3\%$ of all the matter in the universe, with dark matter being $26,8\%$ and baryonic ($i.e.$ ordinary) matter $4,9\%$ \cite{planck}.

Another explanation for the accelerated expansion are alternative theories of gravity, showing that the accelerated expansion might be the door for a generalization of Einstein's theory of gravity. Amongst these proposals, the focus in this work will be the so-called $f(R)$ theories, which will be introduced in the following chapter, and also include nonminimal coupling theories, also developed here.

\cleardoublepage
% %%%%%%%%%%%%%%%%%%%%%%%%%%%%%%%%%%%%%%%%%%%%%%%%%%%%%%%%%%%%%%%%%%%%%%
% Dummy Chapter:
% %%%%%%%%%%%%%%%%%%%%%%%%%%%%%%%%%%%%%%%%%%%%%%%%%%%%%%%%%%%%%%%%%%%%%%

% %%%%%%%%%%%%%%%%%%%%%%%%%%%%%%%%%%%%%%%%%%%%%%%%%%%%%%%%%%%%%%%%%%%%%%
% f(R) theories:
% %%%%%%%%%%%%%%%%%%%%%%%%%%%%%%%%%%%%%%%%%%%%%%%%%%%%%%%%%%%%%%%%%%%%%%
\chapter{\texorpdfstring{$f(R)$}{f(R)} theories}
\label{cap:fofR}

The origins of inflation and its accelerated expansion were first attributed to the presence of a scalar-field. It was this same suggestion that was first applied to the present accelerated expansion of the Universe. Indeed, one of the main mathematical explanations for dark energy other than the Cosmological Constant are scalar fields, like quintessence, $K$-essence and others \cite{darkenergyreview}.

Nonetheless, there are other approaches that may be used to explain the accelerated expansion without recourse to scalar-fiels. Looking at the Einstein-Hilbert action \eqref{eq:einstein-hilbert_action}, the scalar-fields would appear as part of the matter Lagrangian. Another possible explanation would be not to change the matter part but the geometric content of the theory: instead of just resorting to a linear dependence on the Ricci scalar $R$, the possibility would be to use a more evolved function of the latter, $f(R)$. One is led to consider the action

\begin{equation}
S = \int \left[\dfrac{1}{2} f(R) + \LL_m \right] \sqrt{-g} \text{ d}^4 x,
\label{eq:f(R)_action}
\end{equation}

\noindent where the standard Einstein-Hilbert action is recovered if $f(R) = 2\kappa \left(R - 2\Lambda \right)$, with $\kappa=c^4 \slash (16 \pi G)$. Such a modification of the action is not a novelty, for it appears frequently in the studies of Gauss-Bonnet theories, among other proposals. \cite{capozziello1,lovelock}.

This modified action yields the following different field equations through the principle of least action,

\begin{equation}
f_R R_{\mu\nu} - \dfrac{1}{2} f g_{\mu\nu} - \nabla_\mu \nabla_\nu f_R +
\square f_R g_{\mu\nu} = T_{\mu\nu},
\label{eq:field_equations_f(R)}
\end{equation}

\noindent where $f_R \equiv df \slash dR$, $R_{\mu\nu}$ is the Ricci tensor and $\square$ is the D'Alembertian operator $g^{\mu\nu}\nabla_\mu \nabla_\nu$. Like the \ac{EFE} in \eqref{eq:conversation_energy_GR}, these equations also satisfy the conservation of energy $\nabla_\gamma T_{\mu\nu} = 0$ \cite{koivisto}.

Solving these equations for the \ac{FLRW} metric \eqref{eq:metric_flrw} and the perfect fluid energy-momentum tensor \eqref{eq:tensor_perfect_fluid}, the cosmology structure equations of $f(R)$ theories appear as 

\begin{equation}
\begin{split}
3 f_R H^2 =& \dfrac{f_R R - f(R)}{2} - 3 H \dot{f_R} + K^2 \rho, \\
-2 f_R \dot{H} =& \ddot{f_R} - H \dot{f_R} + K^2 \left( \rho + p \right),
\end{split}
\label{eq:cosmology_structure_eq_f(R)}
\end{equation}

\noindent where the dots denote time derivatives.

\section{Scalar-fields equivalence}
\label{sec: scalar-fields equiv}

The theories of $f(R)$ gravity have much more to say in their mathematical aspect, where they end up revealing their equivalence with scalar-fields theory. One of the suggestions starts from the trace of the $f(R)$ field equations \eqref{eq:field_equations_f(R)}, which is 

\begin{equation}
f_R R - 2 f(R) + 3 \square f_R = T,
\label{eq:trace_field_equations_f(R)}
\end{equation}

\noindent with the kinetic term $3\square f_R$ hinting that $f_R$ acts as an additional degree of freedom.

Indeed, in \ac{GR}, the geometrical terms of the trace \eqref{eq:trace_field_equations_f(R)}, reduce to 

\begin{equation}
f(R)=2 k R \Rightarrow f_R = 2k \Rightarrow \square f_R = 0,
\label{eq:condition_gr}
\end{equation}

\noindent and the trace equation becomes simply $R = - T \slash (2k) $, hence $R$ is determined by matter alone.

Let one consider the following action for a scalar-field nonminimally coupled to the curvature,

\begin{equation}
S_\chi = \int \left( \dfrac{1}{2} \left[ f(\chi) + f_\chi (\chi) \left(R - \chi \right) \right] + \LL_M \right) \sqrt{-g} \text{ d}^4 x,
\label{eq:action_scalar_equiv_f(R)}
\end{equation}

\noindent where $\chi$ is a scalar-field and $f_\chi = df/d\chi$. The variation of this action with respect to $\chi$ yields

\begin{equation}
f_{\chi\chi} (\chi) \left( R - \chi \right) = 0,
\label{eq:variation_scalar_resp_to_chi}
\end{equation}

\noindent where $f_{\chi\chi}=df^2 \slash d^2 \chi$. Assuming that $f_{\chi\chi} \neq 0$ (otherwise $f(R)$ is linear and \ac{GR} is recovered), then the only solution for the above equation is $\chi = R$. If this is so, then the action $S_\chi$ reduces to the standard $f(R)$ action $S$ from \eqref{eq:f(R)_action}.

The action $S_\chi$ may also be written as

\begin{equation}
S_\chi = \int \left( 2\kappa \left[ \varphi R - U(\varphi) \right] + \LL_m \right) \sqrt{-g} \text{ d}^4 x,
\label{eq:action_scalar_equiv_f(R)_1}
\end{equation}

\noindent where $\varphi = f_\chi (\chi)$ and the field potential is

\begin{equation}
U(\varphi) = \chi (\varphi) \varphi - f(\chi(\varphi)).
\label{eq:scalar_field_potential}
\end{equation}

The action $S_\chi$ in the form \eqref{eq:action_scalar_equiv_f(R)_1} is important because of its similarity with the \ac{BD} theory of gravity \cite{bransdicke}. This alternative theory to \ac{GR} works with both geometry and a scalar field governing gravitational interactions, rather than only the geometry of spacetime. The action of the \ac{BD} theory with the potential $U(\varphi)$ is,

\begin{equation}
S_{BD} = \int \left( 2\kappa \left[ \varphi R - \dfrac{\omega_{BD}}{ \varphi} \left(\nabla \varphi \right)^2 - U(\varphi) \right] + \LL_m  \right)   \sqrt{-g} \text{ d}^4 x,
\label{eq:action_brans_dicke}
\end{equation}

\noindent where $\omega_{BD}$ is the \ac{BD} parameter and $\left( \nabla \varphi \right) ^2 \equiv \nabla^\mu \varphi \nabla_\mu \varphi$.

If $\omega_{BD} = 0$, then the action $S_{BD}$ becomes equal to the action $S_\chi$ from \eqref{eq:action_scalar_equiv_f(R)_1}, thus showing that $f(R)$ theories can be written as a \ac{BD} theory with a null parameter. As of today, observational results from the Cassini experiment have shown that for \ac{BD} theory to be valid there is the constraint $\omega_{BD}>40000$ \cite{will}. Fortunately, this constraint does not rule out $f(R)$ theories, for it is derived in the absence of a potential $U(\varphi)=0$ in the action and the equivalent action to $f(R)$, in the form \eqref{eq:action_scalar_equiv_f(R)_1}, includes a non-vanishing potential defined in Eq. \eqref{eq:scalar_field_potential}. Hence, the initial consideration of $\omega_{BD}=0$ is not against the quoted constraint.

A possible way of developing further these studies is through conformal transformations to go from the Jordan frame to the Einstein frame: in the former, the scalar field $\varphi$ appears noniminimally coupled to the scalar curvature, thus modyfing the geometrical content of the field dynamics, while a conformal transformation (i.e. a change of units) allows us to transform to the latter, where the scalar curvature appears uncoupled and the scalar field manifests itself as a new matter species, thus allowing for a clearer treatment of the ensuing field equations. For a more developed study on conformal transformations between these two specific frames check Refs. \cite{capozziello2,faraoni1}.

% %%%%%%%%%%%%%%%%%%%%%%%%%%%%%%%%%%%%%%%%%%%%%%%%%%%%%%%%%%%%%%%%%%%%%%

\section{Starobinsky Inflation}
\label{sec:starobinsky}

When first proposed, inflation relied on scalar fields to exist \cite{guth}. However, an inflationary phase with the slow-roll of a scalar field down an appropriate potential can be recast as an $f(R)$ theory, as discussed in the previous paragraph concerning such equivalence.

The $f(R)$ models for inflation are of the following type:

\begin{equation}
f(R) = 2 \kappa \left( R + \alpha R^n \right),
\label{eq:f(R)_inflation_standard}
\end{equation}

\noindent where $\alpha > 0$ and $n > 0$. Considering that the inflation term in the action dominates $f(R) \sim R + \alpha R^n \sim \alpha R^n$ and a De Sitter phase of the universe, such that $T\sim 0$ and $R =\text{const} \Rightarrow \square f_R = 0$, then the trace equation \eqref{eq:trace_field_equations_f(R)} yields

\begin{equation}
\alpha n R^{n-1} R - 2 \alpha R^n = 0 \Leftrightarrow \alpha \left( n-2\right) R^n = 0 \Rightarrow n = 2.
\end{equation}

\noindent This value of $n$ is precisely the case of the first inflationary $f(R)$ model proposed $-$ the Starobinsky model, where $n=2$, written as \cite{starobinsky}:

\begin{equation}
f(R) = 2\kappa \left( R + \dfrac{R^2}{6 M^2} \right),
\label{eq:f(R)_starobinsky}
\end{equation}

\noindent where $M\simeq 3 \times 10^{-6} M_P \sim 10^{13}$ GeV$/c^2$ is a constant with dimensions of mass and $M_P = \sqrt{ (\hbar c) / G } \approx 1.221 \times 10^{19}$ GeV$/c^2 \sim 10^{-8}$ kg is the Planck mass. The value of $M$ relies on data from the \ac{WMAP} normalization of the \ac{CMB} temperature anisotropies \cite{felice,wmap}. The presence of the first linear term of Eq. \eqref{eq:f(R)_starobinsky} will be the reason why inflation will end at a certain point, when it begins to dominate over the quadratic term. For the equivalence of this model $f(R)$ with scalar-field theories, following the analogy of section \ref{sec: scalar-fields equiv}, check Ref. \cite{felice}.

Inserting the above into the cosmology structure equations \eqref{eq:cosmology_structure_eq_f(R)} yields

\begin{eqnarray}
\ddot{H} - \dfrac{\dot{H}^2}{2 H} + \dfrac{1}{2} M^2 H &=& -3 H \dot{H},
\label{eq:cosmology_structure_eq_1_inflation}\\
\ddot{R} + 3 H \dot{R} + M^2 R &=& 0,
\label{eq:cosmology_structure_eq_2_inflation}
\end{eqnarray}

\noindent which are the Starobinsky cosmology equations.

The \ac{SRA} for an inflationary regime implies that $\left| \epsilon \right| = \left| \dot{H}\slash H^2\right| \ll 1$ and $ \left| \ddot{H} \slash \left(H \dot{H}\right) \right| \ll 1$. This means that the first two terms of \eqref{eq:cosmology_structure_eq_1_inflation} can be ignored, so the \ac{SRA} parameter becomes

\begin{equation}
\epsilon \simeq \dfrac{M^2}{6 H^2}.
\label{eq:epsilon_inflation}
\end{equation}

The number of $e$-folds is determined by $N$, which in this case can be written as

\begin{equation}
N \equiv \int^{t_f}_{t_i} H dt \simeq \dfrac{1}{2 \epsilon_1(t_i)},
\end{equation}

\noindent where $t_i$ and $t_f$ are the instants of time on which the inflationary starts and stops.

\subsection{Reheating}
\label{sec:reheating}

The end of inflation happens when the \ac{SRA} parameter becomes of the order $\epsilon \sim 1$, meaning that $H_f \sim M \slash \sqrt{6}$, from Eq. \eqref{eq:epsilon_inflation}. This is followed by a phase of reheating of the Universe, as can be assessed through Eq. \eqref{eq:cosmology_structure_eq_2_inflation}, which with the substitution $R \to a^{-3/2} R$ becomes

\begin{equation}
\ddot{R} + \left( M^2 - \dfrac{3}{4} H^2 - \dfrac{3}{2} \dot{H} \right) R = 0.
\label{eq:cosmology_structure_eq_2_inflation_substitut}
\end{equation}

Reheating is characterized by $M^2 \gg \{ H^2,|\dot{H}| \}$, which leads to two distinct solutions of \eqref{eq:cosmology_structure_eq_2_inflation_substitut}: an harmonic oscillator and a damped oscillation around $R=0$ such as

\begin{equation}
R \propto a^{-3/2} \sin (M t).
\label{eq:damped_oscillation_reheating}
\end{equation}

As for the Hubble parameter, Eq. \eqref{eq:cosmology_structure_eq_1_inflation} is used. Neglecting the $r.h.s.$ of it yields a solution of the type $H(t) \propto \cos^2 (Mt/2)$, meaning that the solution of the complete equation can be written as

\begin{equation}
H(t) = f(t) \cos^2 (Mt/2).
\end{equation}

Considering $R(t) \simeq 6 \dot{H}$ during reheating and making the connection to the slow-roll regime with $\dot{H} = -M^2/6$, the curvature expression becomes (check Ref. \cite{reheating_detail} for details):

\begin{equation}
R(t) \simeq 6 \dot{H} = -3 M f(t) \sin \left[ M (t-t_0) \right],
\end{equation}

\noindent with $t_0$ as the beginning of the reheating phase. The function $f(t)$ ends up being

\begin{equation}
f(t) = \left[ \dfrac{3}{M} + \dfrac{3}{4} (t-t_0) + \dfrac{3}{4M} \sin \left( M \left[t - t_0 \right] \right) \right]^{-1}.
\end{equation}

Notice that after inflation, one has the late-time approximation $M\left(t-t_0\right) \gg 1$, so the previous function becomes

\begin{equation}
f(t) \simeq \dfrac{4}{3 (t-t_0)},
\end{equation}

which means that 

\begin{eqnarray}
H(t) &\simeq & - \dfrac{4}{3(t-t_0)} \cos^2 \left[ \dfrac{M}{2} (t-t_0) \right], \\
R(t) &\simeq & - \dfrac{4M}{t-t_0} \sin \left[M (t-t_0) \right], \label{eq:curvature_reheating_M>1}
\end{eqnarray}

\noindent where $\left< H \right> \simeq (2/3) (t-t_0)^{-1}$ shows that the Universe behaves as if it were matter-dominated during the reheating after Starobinsky inflation.

It was in this phase that most of the matter today on the Universe was formed. This phenomenon is introduced with a scalar field $\chi$ with mass $m_\chi$, together with a nonminimal coupling of the field with the scalar curvature, depicted in the action

\begin{equation}
S = \int d^4x \sqrt{-g} \left[ \dfrac{1}{2} f(R) - \dfrac{1}{2} g^{\mu\nu} \partial_\mu \chi \partial_\nu \chi - \dfrac{1}{2} m^2_\chi \chi^2 - \dfrac{1}{2} \xi R \chi^2 \right],
\label{eq:action_preheating_starobinsky}
\end{equation}

\noindent where $f(R)$ is the quadratic function defined in \eqref{eq:f(R)_starobinsky}.

The variation of this action with respect to $\chi$ yields

\begin{equation}
\square \chi - m^2_\chi \chi - \xi R \chi = 0.
\label{eq:field_equation_preheating}
\end{equation}

As $\chi$ is a quantum field, it can be decomposed into modes according to

\begin{equation}
\chi (t, \textbf{x}) = \dfrac{1}{(2\pi)^{3/2}} \int d^3k \left[a_k \chi_k (t) e^{-i \textbf{k}\cdot\textbf{x}} + a^\dagger_k \chi^\star_k (t) e^{i \textbf{k}\cdot\textbf{x}} \right],
\end{equation}

\noindent where $a^\dagger_k$ and $a_k$ are creation and annihilation operators of particles with mass $m_\chi$ and momentum $\textbf{k}$, from quantum field theory. Each Fourier mode $\chi_k (t)$ is governed by the equation of motion

\begin{equation}
\ddot{\chi}_k + 3 H \dot{\chi}_k + \left( \dfrac{k^2}{a^2} + m^2_\chi + \xi R \right) \chi_k = 0,
\label{eq:fourier_mode_preheating}
\end{equation}

\noindent where $k = | \textbf{k}|$ is a comoving wavenumber. Performing the substitution $\chi_k \to a^{-3/2} \chi_k$ transforms the last equation into

\begin{equation}
\ddot{\chi}_k + \left( \dfrac{k^2}{a^2} + m^2_\chi + \xi R - \dfrac{9}{4} H^2 - \dfrac{3}{2} \dot{H} \right) \chi_k = 0.
\end{equation}

The case where $|\xi| > 1$ allows to neglect the last two terms inside the bracket of this equation. Moreover, using the approximated expression for the curvature from \eqref{eq:curvature_reheating_M>1}, the above equation becomes

\begin{equation}
\ddot{\chi}_k + \left[\dfrac{k^2}{a^2} + m_\chi^2 - \dfrac{4 M \xi}{t-t_0} \sin { M(t-t_0) } \right] \chi_k \simeq 0.
\end{equation}

\noindent If the term inside brackets is defined as

\begin{equation}
\omega^2_k \equiv \dfrac{k^2}{a^2} + m_\chi^2 - \dfrac{4 M \xi}{t-t_0} \sin { M(t-t_0) },
\label{eq:frequency_reheating}
\end{equation}

\noindent then the equation for $\chi_k$ in this regime can be written as simply as

\begin{equation}
\ddot{\chi}_k + \omega_k^2 \chi_k \simeq 0.
\end{equation}

Particle production is thus obtained with the resonance of this oscillator --- a phenomenon of parametric resonance which can be more detailed by defining $z$ with $M(t-t_0) = 2z \pm \pi / 2$, where the opposite signs are related to the sign of $\xi$. The new variable leads to what is called the Mathieu equation

\begin{equation}
\dfrac{d^2 \chi_k}{dz^2} + \left[A_k - 2 q \cos (2z) \right] \chi_k \simeq 0,
\label{eq:mathieu_eq_starobinsky}
\end{equation}

\noindent with $A_k$ and $q$ determining the strength of parametric resonance and defined by

\begin{equation}
A_k = \dfrac{4 k^2}{a^2 M^2} + \dfrac{4 m_\chi^2}{M^2},~~~~~~ q=\dfrac{8|\xi|}{M(t-t_0)}.
\end{equation}

\begin{figure}[ht]
\centering
\includegraphics[width=0.5\textwidth]{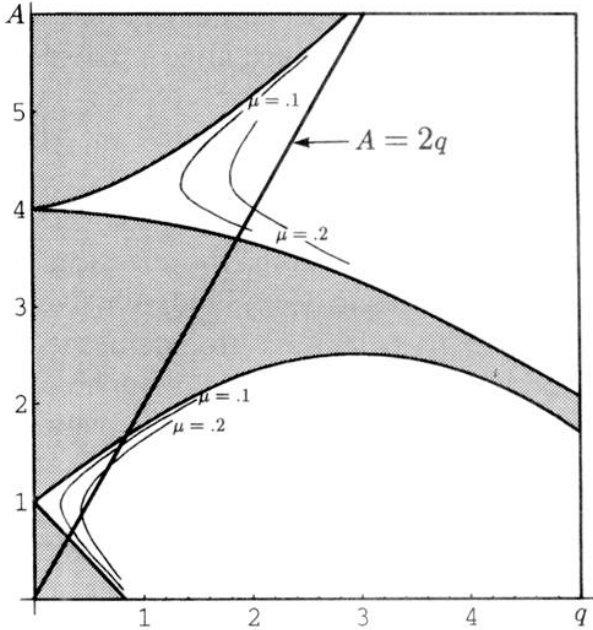}
\caption{Sketch of the stability/instability chart of the canonical Mathieu equation \cite{mathieu_map1}.}
\label{fig:floquet chart}
\end{figure}

The effect of the parametric resonance is obtained through a stability-instability map proper of the Mathieu equation, shown in figure \ref{fig:floquet chart}. The white bands of the chart are the regions of instability, the grey bands correspond to regions of stability and the curve lines inside the white bands show the values of an instability parameter $\mu_k$, which yields the strength of the exponential growth of the instability of each mode $\chi_k$. The line $A=2q$ corresponds to the values of $A$ and $q$ for $\cos(2z)=1$. The oscillatory aspect of the field implies that the resonance growth changes with the expansion of the Universe, which also causes the parameters $A_k$ and $q$ to depend on time.

This process of particle production due to parametric resonance of the field is called preheating. It should still be said this is not necessary for particle production to occur. Indeed, if $\xi=0$, which is the standard reheating regime, there still is a parametric oscillation with $q=0$ due to the expansion of the Universe alone. Nonetheless, the preheating scenario is much more efficient than standard reheating.

% %%%%%%%%%%%%%%%%%%%%%%%%%%%%%%%%%%%%%%%%%%%%%%%%%%%%%%%%%%%%%%%%%%%%%%

\section{Solar System Effect}
\label{sec:solar_sys_impact}

Even though $f(R)$ gravity might explain the accelerated expansion of the Universe in the inflation and present eras, its validity at the Solar System scale should be assessed. The long-range impact of $f(R)$ in the Solar System is taken into account by considering the flat \ac{FLRW} metric with a spherically symmetric perturbation

\begin{equation}
ds^2 = - \left[1 + 2\Psi (r) \right] dt^2 + a(t)^2 \left( \left[1 + 2 \Phi(r) \right] dr^2 + r^2 d\Omega^2 \right),
\label{eq:metric_chiba}
\end{equation}

\noindent where $\Psi$ and $\Phi$ are perturbing functions such that $\left|\Psi(r)\right| \ll 1$ and $\left|\Phi(r)\right| \ll 1$.

The background curvature is nonzero, $R_0 (t) \neq 0$, thus it can be decomposed into a sum of two components:

\begin{equation}
R(r,t) \equiv R_0 (t) + R_1 (r),
\label{eq:curvature_chiba}
\end{equation}

\noindent where $R_1(r)$ is considered as a time-independent perturbation of the background curvature. This analysis follows the work of Ref. \cite{chiba}. The validity of the following treatment assumes that $R_1 \ll R_0$.

The background curvature solves the trace of the field equations \eqref{eq:trace_field_equations_f(R)} according to

\begin{equation}
f_{R0} (t) R_0 (t) - 2 f_0 (t) + 3\square f_{R0} (t) = T^\text{cos} (t),
\label{eq:trace_chiba_background}
\end{equation}

\noindent where $f_{R0} \equiv df \slash dR \big|_{R=R_0}$, $f_0 \equiv f(R_0)$ and $T^\text{cos}$ is the trace of the energy-momentum tensor of cosmological baryonic matter.

Considering that all derivatives of $f(R)$ are well defined at $R_0 (t_0)$, where $t_0$ is the present time, a Taylor expansion around $R_0$ can be used to write the functions $f(R_0 + R_1)$ and $f_R (R_0 + R_1)$. As the framework is a weak-field regime, the expansion neglects terms non-linear in the perturbation $R_1$, provided that the following conditions are taken into account:

\begin{eqnarray}
f_0 + f_{R0} R_1 &\gg &\dfrac{1}{n!} f^{(n)} (R_0) R^n_1, \nonumber \\
f_{R0} + f_{RR0} R_1 &\gg &\dfrac{1}{n!} f^{(n+1)}(R_0) R^n_1 \text{ for all } n>1,
\label{eq:conditions_linearization_chiba}
\end{eqnarray}

\noindent where $f_{RR0} \equiv d^2 f \slash dR^2 \big|_{R=R_0}$ and $f^{(n)} (R_0) = d^n f \slash dR^n \big|_{R=R_0}$.

Applying these expansions into the complete trace equation

\begin{equation}
f_R R - 2 f + 3 \square f_R =  T^\text{cos} + T^\text{s},
\label{eq:trace_chiba_complete}
\end{equation}

\noindent where $T^\text{s}$ is the energy-momentum tensor of a spherically symmetric mass source, the trace equation becomes linearized as

\begin{equation}
3 f_{RR0} \square R_1(r) - \left[ f_{R0} (t) - f_{RR0} (t) R_0 (t) - 3\square f_{RR0} (t) \right] R_1 = T^\text{s},
\label{eq:trace_chiba_complete_linearized}
\end{equation}

\noindent where equation \eqref{eq:trace_chiba_background} was used to eliminate all the terms independent of $R_1$. It is worthy of note that this linearization is taking into account the already stated condition $R_1 \ll R_0 \neq 0$, due to the presence of the term $f_{RR0} R_0 R_1$, while neglecting terms of the order $f_{RR0} R_1^2$.

Since $R_1(r)$ is time-independent, the operator $\square$ becomes $\square \approx \nabla^2$. The matter at the local perturbation will be considered with no pressure, such that $T^\text{s} = -\rho (r) c^2$. Therefore, the equation \eqref{eq:trace_chiba_complete_linearized} can be written as 

\begin{equation}
\nabla^2 R_1 - m^2 R_1 = - \dfrac{\rho}{3 f_{RR0}},
\label{eq:trace_equation_chiba_final}
\end{equation}

\noindent with

\begin{equation}
m^2\equiv  \frac{1}{3} \left( \frac{f_{R0}}{f_{RR0}} - R_0 - 3 \frac{\square f_{RR0}}{f_{RR0}} \right).
\label{eq:m_parameter}
\end{equation}

\noindent This equation can be solved by means of the Green function

\begin{equation}
G(r) = \left\{
  \begin{array}{l l}
     - \cos (mr) \slash (4 \pi r) & \quad \text{if  } m^2 < 0\\
     \quad \\
     - \exp (-mr) \slash (4 \pi r) & \quad \text{if  } m^2 > 0   \end{array} \right. .
\label{eq:green_function_chiba}
\end{equation}

\subsection{Long Range Regime}
\label{subsec:long-range}

Considering first the long range regime $mr \ll 1$, both parts of the Green function can be approximated by $-1 \slash (4\pi r)$. This means that the term linear in $R_1$ from \eqref{eq:trace_equation_chiba_final} may disappear and the solution for the curvature perturbation becomes

\begin{equation}
R_1 = \dfrac{1}{12 \pi f_{RR0}}\dfrac{M_S}{r},
\label{eq:curvature_solution_chiba}
\end{equation}

\noindent where $M_S$ is the total mass of the solar-like source.

Adopting the arguments for linearization from \eqref{eq:conditions_linearization_chiba}, using the background curvature equation \eqref{eq:trace_chiba_complete_linearized} and neglecting all the terms that are not linear in $R_1$, $\Phi$ and $\Psi$, the solutions for these perturbations are taken from the time ($tt$) and radial ($rr$) components of the field equations.

The linearization of the $tt$ component of \eqref{eq:field_equations_f(R)} has the form

\begin{equation}
f_{R0} \nabla^2 \Psi + \dfrac{1}{2} f_{R0} R_1 - f_{RR0} \nabla^2 R_1 = \rho.
\label{eq:tt_component_FE_chiba}
\end{equation}

Importing equation \eqref{eq:trace_equation_chiba_final} with $m^2=0$ reduces the above to

\begin{equation}
f_{R0} \nabla^2 \Psi = \dfrac{2}{3} \rho - \dfrac{1}{2} f_{R0} R_1,
\end{equation}

\noindent which can be decomposed according to $\Psi = \Psi_0 + \Psi_1$, where

\begin{eqnarray}
f_{R0} \nabla^2 \Psi_0 &=& \dfrac{2}{3} \rho, \label{eq:psi_0_eq_chiba}\\
f_{R0} \nabla^2 \Psi_1 &=& - \dfrac{1}{2} f_{R0} R_1. \label{eq:psi_1_eq_chiba}
\end{eqnarray}

The first of these is integrated with Gauss's Theorem and becomes

\begin{equation}
\Psi_0'(r) = \dfrac{1}{6 \pi f_{R0}} \dfrac{m(r)}{r^2},
\label{eq:psi_0_'_eq_chiba}
\end{equation}

\noindent where the prime represents differentiation with respect to $r$ and $m(r)$ is the mass of a sphere with radius $r$. Assuming that $\lim_{r \to \infty} \Psi_0 = 0$, the solution for $\Psi_0$ is

\begin{equation}
\Psi_0 = - \dfrac{1}{6\pi f_{R0}}\dfrac{M_S}{r}.
\end{equation}

One the other hand, using the solution for $R_1$ from \eqref{eq:curvature_solution_chiba} in \eqref{eq:psi_1_eq_chiba} and integrating it outside the star, gives

\begin{equation}
\left| \Psi_1 \right| = \dfrac{1}{48 \pi f_{RR0}} M_S r.
\end{equation}

To treat this equation it is relevant to expand the condition $mr\ll 1$ with the definition of $m$, presented after Eq. \eqref{eq:trace_equation_chiba_final}:

\begin{equation}
\begin{split}
\left| m^2 \right| r^2 \equiv \left| \dfrac{1}{3} \left( \dfrac{f_{R0}}{f_{RR0}} - R_0 - 3 \dfrac{\square f_{RR0}}{f_{RR0}}\right) \right| r^2 &\ll 1 \\
\Rightarrow \left| \dfrac{f_{R0}}{f_{RR0}}\right| r^2 - \left| R_0 - 3 \dfrac{\square f_{RR0}}{f_{RR0}} \right| r^2 &\ll 1,
\end{split}
\label{eq:mr_condition_expanded}
\end{equation}

\noindent where the triangle inequality was used. Taking into account that $\square f_{RR0} \slash f_{RR0} \sim H^2$ and knowing by cosmological constraints that $R_0 r^2 \sim H_0^2 r_0^2 \ll 1$, then the constraint $mr \ll 1$ can be written as

\begin{equation}
\left|\dfrac{f_{R0}}{f_{RR0}}\right| r^2 \ll 1.
\label{eq:mr_condition_as_f(R)}
\end{equation}

This allows to write the useful condition $|\Psi_1| \sim M_S r \slash f_{RR0} \ll M_S \slash (f_{R0} r) $. Since $\Psi_0 \sim M_S \slash (f_{R0} r)$ implies the condition $\left| \Psi_1 \right| \ll \left| \Psi_0 \right|$, then $\Psi_1$ may be neglected, thus $\Psi \equiv \Psi_0$.

Following the same procedures, the linearization of the $rr$ component of \eqref{eq:field_equations_f(R)} takes the form

\begin{equation}
f_{R0} \left( - \Psi'' + \dfrac{2}{r} \Phi' \right) - \dfrac{1}{2} f_{R0} R_1 + \dfrac{2}{r} f_{RR0} R_1' = 0.
\label{eq:rr_component_FE_chiba}
\end{equation}

First of all, the second term of the equation can be neglected with respect to the last one, due to the application of \eqref{eq:mr_condition_as_f(R)} as

\begin{equation}
\left| \dfrac{(1/2) f_{R0} R_1}{2 f_{RR0} R_1' / r} \right| \sim \left| \dfrac{f_{R0}}{f_{RR0}} \right| r^2 \ll 1.
\end{equation}

Computing $\Psi''$ from \eqref{eq:psi_0_'_eq_chiba} and $R_1'$ from \eqref{eq:trace_equation_chiba_final} (with $m^2 = 0$), Eq. \eqref{eq:rr_component_FE_chiba} can be written as

\begin{equation}
\Phi'(r) = \dfrac{1}{12 \pi f_{R0}} \dfrac{d}{dr} \left[ \dfrac{m(r)}{r}\right],
\label{eq:phi'_eq_chiba}
\end{equation}

which can be integrated outside the matter source and with the outer limit $\lim_{r \to \infty} \Phi = 0$ to give

\begin{equation}
\Phi = \dfrac{1}{12 \pi f_{R0}}\dfrac{M_S}{r}.
\label{eq:phi_eq_chiba}
\end{equation}

Comparing the solution $\Phi$ with $\Psi$, from \eqref{eq:psi_0_eq_chiba}, it is easy to conclude that $\Psi = -2 \Phi$, thus implying that the \ac{PPN} parameter $\gamma = - \Phi \slash \Psi $ takes the value $\gamma = 1/2$, which contradicts the observational results that yield $\gamma \sim 1$ \cite{will} --- implying that, if the additional degree of freedom is long-ranged ($mr \ll 1$), $f(R)$ gravity is incompatible with experiment.
Notice that, in \ac{BD} theories, $\gamma = 1/2$ is obtained from $\omega = 0$: following the discussion after Eq. \eqref{eq:action_brans_dicke}, the very light mass implies that the potential has no impact on the dynamics, and as such this result only restates that an $f(R)$ model with this characteristic is observationally excluded.

\subsection{Short Range Regime}
\label{subsec:short-range}

To obtain a valid $f(R)$ model one should then study the case where the spherical symmetric mass is sufficiently big to make an impact, \textit{ie} the short range regime of $f(R)$ should be studied. And, to study the short range regime of $f(R)$ gravity in the Solar System, it is more adequate to consider a perturbation of a Minkowski metric of the form

\begin{equation} \label{eq:metric_naf}
ds^2=-\left[1+2\Psi\left(r\right)\right]c^2dt^2+\left[1+2\Phi(r)\right] dr^2 + r^2 d\Omega^2,
\end{equation}

\noindent where $\Psi$ and $\Phi$ are again perturbing functions such that $\left|\Psi(r)\right| \ll 1$ and $\left|\Phi(r)\right| \ll 1$. This is so because if $f(R)$ is short-ranged ($mr\gg 1$), the dynamics of the Universe do not have any impact at Solar System scales, while in the opposite case (long range, $mr\ll 1$), the effect of $f(R)$ extends to much longer distances, so that the asymptotic values of cosmology must be taken into account.

In the following it is assumed that matter behaves as dust, $i.e.$ a perfect fluid with negligible pressure and an energy-momentum tensor described by
\begin{equation}
T_{\mu\nu} = \rho c^2 u_\mu u_\nu~~~~,~~~~  u_\mu u^\mu =-1,
\label{eq:conditions_energy_mom_chap2}
\end{equation}
\noindent where $\rho$ is the matter density and $u_\mu$ is the four-velocity vector. The same procedure was followed in the long-range situation, though it was not as detailed as in this section. The trace of the energy-momentum tensor is $T = -\rho c^2$. The Lagrangian density of matter is considered as $\LL_m = -\rho c^2$ (see Ref. \cite{dynimpac1} for a discussion).

The function $\rho = \rho(r)$ is that of a spherically symmetric body with a static radial mass density $\rho = \rho(r)$ and the function $\rho(r)$ and its first derivative are assumed to be continuous across the surface of the body,
\begin{equation} \label{conditions for fe_march}
\rho(R_S) = 0 \quad \text{and} \quad \dfrac{d\rho}{dr}(R_S) = 0,
\end{equation}
\noindent where $R_S$ denotes the radius of the spherical body. These conditions play a role in the following sections, where integrals that have $R_S$ as an integration limit appear.

This analytical setup allows to define the function $f(R)$ as a Taylor expansion around $R=0$, according to Ref. \cite{naff}:

\begin{equation}
f(R) \approx 2 \kappa \left( R + \dfrac{R^2}{6 m^2} \right) + \OO (R^3),
\label{eq:f(R)_naff}
\end{equation}

\noindent which is similar to the Starobinsky action functional from \eqref{eq:f(R)_starobinsky}, although here the mass $m$ is related to the second derivative of the more general $f(R)$ function, while in the latter this was considered to be exact.

Repeating what was done in the long range regime, the equations for the curvature and for the perturbations $\Psi$ and $\Phi$ will be linearized. In this case, the linearization is done according to the expansion of each of these three quantities in powers of $c^{-1}$ and all the terms of the order $\OO(c^{-3})$ or less are neglected.

Inserting \eqref{eq:f(R)_naff} into the curvature equation \eqref{eq:trace_field_equations_f(R)} and linearizing it, yields

\begin{equation}
\nabla^2 R - m^2 R = - \dfrac{8 \pi G}{c^2} m^2 \rho,
\label{eq:trace_eq_naff}
\end{equation}

\noindent which through the variable substitution $u=rR$ becomes 

\begin{equation}
\dfrac{d^2 u}{dr^2} - m^2 u = s(r),
\label{eq:trace_eq_subst_naff}
\end{equation}

\noindent where $s(r)= - \left( 8\pi G \slash c^2 \right) m^2 r \rho$ is the source function. The function $u(r)$ respects the following boundary conditions:

\begin{itemize}
\item $u(0)=0$, so that the curvature $R$ may have a finite value at that point;
\item $\lim_{r \to \infty} u(r)=0$, so that the curvature vanishes asymptotically as one recovers the Minkowski metric.
\end{itemize}

The Green function of \eqref{eq:trace_eq_subst_naff} in the sense of distributions solves the equation

\begin{equation}
\dfrac{d^2 G(r,r')}{dr'^2} - m^2 G(r,r') = \delta (r - r'),
\end{equation}

\noindent where $\delta(r-r')$ is a Dirac delta distribution. This Green function $G(r,r')$ is used to determine the solution of the curvature equation by means of the integral $u(r) = \int^r_0 G(r,r') s(r') dr'$.
Due to the different boundary conditions, the curvature is written as a twofold solution:

\begin{equation} \label{curvature solution}
R(r) = \left\{ 
  \begin{array}{l l}
     R^\uparrow(r) & \quad \text{if  } r>R_S\\
     \quad \\
     R^\downarrow(r) & \quad \text{if  } 0\leq r\leq R_S   \end{array} \right. ,
\end{equation}

\noindent where $R^\downarrow(r)$ is the curvature inside the star 

\begin{equation}
R^\downarrow (r) = -\dfrac{4 \pi G}{c^2 m} \left[ \dfrac{e^{-mr}}{r} I_1(r)
+ \dfrac{2 \sinh(mr)}{r} I_2(r) \right].
\end{equation}

The quantities $I_1(r)$ and $I_2(r)$ have been computed by using the properties \eqref{conditions for fe_march} of the mass density $\rho(r)$ and are given by

\begin{eqnarray}
\nonumber I_1(r)&=& - 2 m^2\int^{r}_{0}\sinh(mr')r'\rho(r')dr',\\ \nonumber
I_2(r)&=& - m^2  \int^{R_S}_r e^{-mr'}r'\rho(r')  dr'  .
\end{eqnarray}

\noindent On the other hand, $R^\uparrow(r)$ is the curvature outside the star, given by

\begin{equation} \label{curvature solution, outside star}
R^\uparrow (r) = \dfrac{2 G M_S}{r c^2} m^2 A(m,R_S) e^{-mr},
\end{equation}

\noindent where $A(m,R_S)$ is a form factor defined as

\begin{equation}
A(m,R_S) = \dfrac{4\pi}{m M_S} \int^{R_S}_0 \sinh(mr) r \rho(r) dr,
\end{equation}

\noindent which shall be discussed in a subsequent section.

As for the solution of $\Psi$, it is obtained through the linearization of the $tt$ component of the field equations \eqref{eq:field_equations_f(R)}, neglecting all terms smaller than $\OO(1/c^2)$, combined with the $tt$ component of the Ricci tensor at order $\OO(1/c^2)$,

\begin{equation}
R_{tt} = \dfrac{2\Psi'}{r}+\Psi'' + \OO\left(\dfrac{1}{c^3}\right) = \nabla^2\Psi + \OO\left(\dfrac{1}{c^3}\right),
\label{eq:ricci_tensor_tt_naff}
\end{equation}

\noindent while the curvature equation \eqref{eq:trace_eq_naff} yields the expression ruling $\Psi$:

\begin{equation} \label{eq:psi_equation_naff}
\nabla^2\Psi = \dfrac{c^2}{3k}\rho-
\dfrac{R}{6}.
\end{equation}

This is solved outside the spherical body, $r \geq R_S$, by decomposing it into a sum,
\begin{equation} \label{psi equation, decomposition}
\nabla^2 \Psi = \nabla^2 \Psi_0 + \nabla^2 \Psi_1,
\end{equation}
\noindent where
\begin{equation} \label{psi 1 and psi 2}
\nabla^2 \Psi_0 = \dfrac{c^2}{3k} \rho~~~~,~~~~\nabla^2 \Psi_1 = -\dfrac{R}{6}.
\end{equation}
The first equation is easily solved using the divergence theorem, which gives
\begin{equation} \label{psi 0}
\Psi_0 (r) = - \dfrac{4 }{3 } \dfrac{GM_S}{c^2r}.
\end{equation}
The second one needs a longer integration, which eventually shows that
\begin{equation} \label{psi 1}
\Psi_1 (r) = \dfrac{G M_S}{3 c^2 r} \left[ 1 - A(m,R_S) e^{-mr} \right],
\end{equation}
\noindent so that for $r \geq R_S$ the solution is
\begin{equation} \label{psi solution}
\Psi(r) = - \dfrac{G M_S}{c^2 r} \left[1 + \dfrac{1}{3} A(m,R_S) e^{-mr}\right].
\end{equation}

Taking the $rr$ component of the Ricci tensor to the usual order of $\OO(1/c^2)$ and recording that the mass distribution is static yields

\begin{equation} 
R_{rr} = \dfrac{2}{r}\Phi' - \Psi'' + \OO\left(\dfrac{1}{c^3}\right)~~~~,~~~~
T_{rr} = 0,
\label{eq:radial_components_naff}
\end{equation}

\noindent so that the $rr$ component of the field equations \eqref{eq:field_equations_f(R)} becomes

\begin{equation} \label{eq:phi_equation_naff}
2\Phi' - r \Psi'' - \dfrac{r R}{2} + \dfrac{2 R'}{3 m^2} = 0.
\end{equation}
\noindent This equation is easily integrated outside the spherical body, $r \geq R_S$, leading to
\begin{equation} \label{phi solution}
\Phi(r)=\dfrac{G M_S}{c^2 r} \left[ 1 - \dfrac{1}{3} A(m,R_S) e^{-mr} (1 +mr) \right].
\end{equation}

\noindent In the GR limit, $m \to \infty$, the exponential term in both $\Psi$ and $\Phi$ vanishes and the weak-field approximation of the Schwarzschild metric is obtained, as expected from a weak spherical perturbation on a Minkowski metric.

It should be noted that if the metric \eqref{eq:metric_naf} is written in the isotropic coordinates

\begin{equation}
ds^2 = - \left[ 1 + 2 \Psi' (r') \right] dt'^2 + \left[ 1 + 2 \Phi' (r') \right] \left(dr'^2 + r'^2 d\Omega'^2 \right],
\label{eq:metric_isotropic_naff}
\end{equation}

then the perturbing functions \eqref{psi solution} and \eqref{phi solution} become

\begin{eqnarray}
\Psi' (r') &=& - \dfrac{G M_S}{c^2 r'} \left[1 + \dfrac{1}{3} A(m,R_S) e^{-mr} \right], \\
\Phi' (r') &=& \dfrac{G M_S}{c^2 r'} \left[1 - \dfrac{1}{3} A(m,R_S) e^{-mr} \right],
\end{eqnarray}

\noindent which differ only in the signal of the second term.

These results should not be analysed with the \ac{PPN} formalism, as they instead depict a Yukawa type correction to the Newtonian potential with a field strength $1/3$. This kind of Yukawa correction to the Newtonian potential is not new in physics and many experimental tests have been made on it, as shall be discussed in a subsequent chapter.

\cleardoublepage

% %%%%%%%%%%%%%%%%%%%%%%%%%%%%%%%%%%%%%%%%%%%%%%%%%%%%%%%%%%%%%%%%%%%%%%
% Dummy Chapter:
% %%%%%%%%%%%%%%%%%%%%%%%%%%%%%%%%%%%%%%%%%%%%%%%%%%%%%%%%%%%%%%%%%%%%%%

% %%%%%%%%%%%%%%%%%%%%%%%%%%%%%%%%%%%%%%%%%%%%%%%%%%%%%%%%%%%%%%%%%%%%%%
% f(R) theories:
% %%%%%%%%%%%%%%%%%%%%%%%%%%%%%%%%%%%%%%%%%%%%%%%%%%%%%%%%%%%%%%%%%%%%%%
\chapter{Nonminimal Coupling}
\label{cap:nmc}

The remarkable range of applications of $f(R)$ gravity, and in particular the Starobinsky inflationary model, prompted physicists to generalize these theories even further. This generalization is made by a \ac{NMC} between geometry and matter, through the product of the matter Lagrangian density with another function of the curvature, as posited by the action \cite{NMCmodel}

\begin{equation}\label{eq:action_nmc}
S_\text{NMC}=\int \left[\dfrac{1}{2} f^1(R)+\left[1+f^2(R)\right] \LL_m\right]
\sqrt{-g} d^4x.
\end{equation}

\noindent The field equations, obtained through the variation of the action with respect to the metric, are

\begin{equation} \label{eq:field_equations_nmc}
\left(f^1_R + 2 f^2_R \LL_m \right) R_{\mu \nu} -
\dfrac{1}{2} f^1 g_{\mu \nu} =
\left(1+ f^2 \right) T_{\mu \nu} +\left( \square_{\mu \nu} - g_{\mu \nu} \square \right)
\left( f^1_R + 2 f^2_R \LL_m \right),
\end{equation}

\noindent where $f_R^i \equiv df^i (R) /dR$. If $f^1(R) = 2 \kappa ( R - 2\Lambda)$ and $f^2(R) = 0$, then the action \eqref{eq:action_nmc} collapses to the Einstein-Hilbert action from \eqref{eq:einstein-hilbert_action}. The Palatini formulation of a NMC theory, obtained by considering that the metric and affine connection are {\it a priori} independent, can be found in Ref. \cite{PalatiniNMC}.

It should be said that, in opposition to \ac{GR} and pure $f(R)$ gravity, these field equations do not respect the conservation of energy law, $\nabla^\mu T_{\mu\nu} \neq 0$. Indeed, the covariant derivative of the field equations \eqref{eq:field_equations_nmc}, together with the Bianchi identities $\nabla^\mu G_{\mu\nu} = 0$ and the identity

\begin{equation}
(\square \nabla_\nu - \nabla_\nu \square ) f_R^i = R_{\mu\nu} \nabla^\mu f_R^i,
\end{equation}

\noindent yields for the covariant derivative of the energy-momentum tensor,

\begin{equation}
\nabla^\mu T_{\mu\nu} = \dfrac{f_R^2}{1+f^2} \left(g_{\mu\nu} \LL_m - T_{\mu\nu} \right) \nabla^\mu R.
\end{equation}

\noindent Nonetheless, if $f^2(R) = 0$, the pure $f(R)$ case is recovered and the conservation law is fulfilled. This non-conservation of the energy-momentum tensor may yield a violation of the Equivalence principle, for the nonminimal coupling between geometry and matter means that the motion of matter is dependent on the specific form of the Lagrangian density $\LL_m$ and thus matter may not follow the expected geodesic motion.

It should also be mentioned that \ac{NMC} has an impact on the description of the interior of a central body, leading to a correction to the latter's coupling strength, as is illustrated in Refs. \cite{stelobserv,mimlambda,dynimpac2}.

\section{Scalar-fields equivalence}
\label{sec:scalar-fields}

In section \ref{sec: scalar-fields equiv}, the pure $f(R)$ gravity theories were shown to be equivalent to scalar-tensor theories $-$ in particular, to a \ac{BD} theory of gravity. \ac{NMC} theories of gravity can also be written in a similar way, though with two independent scalar fields \cite{damour}. These theories are introduced through the action

\begin{equation}
S = \int \left[ \dfrac{1}{2} f^1 (\phi) + \left[ 1 + f^2 (\phi) \right] \LL_m + \psi (R - \phi) \right] \sqrt{-g} d^4 x,
\label{eq:action_scalar_field_nmc}
\end{equation}

\noindent where $\psi$ acts as a Lagrange multiplier imposing the relation $\phi = R$, with which the action $S$ is reduced to $S_{NMC}$. The variation of this action \eqref{eq:action_scalar_field_nmc} with respect to $\phi$ then yields

\begin{equation}
\psi = \dfrac{1}{2} f^1_R (\phi) + f_R^2 (\phi) \LL_m,
\end{equation}

\noindent which indicates that both scalar-fields are independent if $\LL_m \neq 0$ or $f_R^2 \neq 0$. If $f_R^2=0$, then the action returns to the pure $f(R)$ case.

As in the pure $f(R)$ situation, the action \eqref{eq:action_scalar_field_nmc} can be written in the form of a \ac{BD} theory with $\omega_{BD} = 0$ (check Eq. \eqref{eq:action_brans_dicke}) and the same scalar-fields,

\begin{equation}
S = \int \left[ \psi R - V(\phi,\psi) + \left[ 1 + f^2(\phi) \right] \LL_m \right] \sqrt{-g} d^4x,
\end{equation}

\noindent with the potential

\begin{equation}
V(\phi,\psi) = \phi \psi - \dfrac{1}{2} f^1 (\phi).
\end{equation}

% %%%%%%%%%%%%%%%%%%%%%%%%%%%%%%%%%%%%%%%%%%%%%%%%%%%%%%%%%%%%%%%%%%%%%%

\section{Preheating and Inflation}
\label{sec:preheating}

In section \ref{sec:reheating}, the process of particle production after inflation was explained by means of a preheating mechanism, whose main characteristic was on the parametric resonance of a quantum field $\chi$. The action of this field included a \ac{NMC} term between the curvature and the field itself, of the type $\xi R \chi^2$ $-$ \textit{i.e.} an additional mass term. This linear \ac{NMC} prompts for the more general assumption of the action \eqref{eq:action_nmc} with the functions

\begin{equation}
f^1(R) = 2\kappa \left( R + \dfrac{R^2}{6 M^2} \right), ~~~\qquad\qquad~~~~~~ f^2(R) = 2\xi \dfrac{R}{M^2}.
\end{equation}

\noindent Logically, this specific action has to be able to explain inflation at the standard \ac{SRA}, which means that $f^2(R)$ should interfere only in the preheating situation, and so the \ac{NMC} model only works as part of a perturbative regime with $f^2(R) \sim 0$.

As stated in section \ref{sec:starobinsky}, the mass parameter has the value $M \sim 10^{13}$ GeV$/c^2$, whereas the \ac{NMC} parameter $\xi$ is dimensionless and subject to the range $1 < \xi < 10^4$. The lower bound $\xi > 1$ was established in order to provide the validity of the condition $R \gg M^2 \slash (2\xi)$ throughout the \ac{SRA} regime. The upper bound of $\xi < 10^4$ is obtained by resorting to the initial inflationary temperature \cite{reheating}.

As in Eq. \eqref{eq:fourier_mode_preheating}, the decomposition of the scalar field from the \ac{NMC} in Fourier modes yields that each mode is governed by

\begin{equation}
\ddot{\chi}_k + \left( 3H + \dfrac{f^2_R}{\left(1 + f^2\right)} \dot{R} \right) \dot{\chi}_k + \left( \dfrac{k^2}{a^2} + m^2\right) \chi_k = 0.
\end{equation}

\noindent The redefinition $X_k \equiv a^{3/2} \sqrt{f^2} \chi_k$ and the new variable $z$ defined by $2 z = M \left(t - t_0 \right) \pm \pi / 2$ (depending on the sign of $\xi$) allows to transform the previous equation into

\begin{equation}
X_k'' + \left[\left(\dfrac{2k}{aM}\right)^2 + \left(\dfrac{2m}{M}\right)^2 - 3 \dfrac{H'}{M}-9 \dfrac{H^2}{M^2} + \dfrac{1}{2}\dfrac{f_R^2}{\left( 1 + f^2\right)}\left(\dfrac{1}{2}\dfrac{f_R^2}{\left( 1 + f^2\right)} R'^2 - 6 \dfrac{HR'}{M} - R'' \right) \right] X_k = 0,
\end{equation}

\noindent where the prime represents a derivative with respect to $z$. Recalling that this action is valid in a $f^2(R) \approx 0$ regime, the above equation can be expanded as,

\begin{equation}
X_k'' + \left[\left(\dfrac{2k}{aM}\right)^2 + \left(\dfrac{2m}{M}\right)^2 - 3 \dfrac{H'}{M}-9 \dfrac{H^2}{M^2} + + \dfrac{\xi}{M^2} \left( \xi \dfrac{R'^2}{M^2} - 6 \dfrac{H R'}{M} - R'' \right) \right] X_k = 0.
\end{equation}

As discussed before, the fact that after the \ac{SRA} the parameter $z$ answers to the late-time approximation $z \gg 1$ makes the Hubble parameter and curvature go through an oscillatory phase of the type

\begin{equation}
H(z) \simeq \dfrac{M}{3z} \left[1 + \sin (2z) \right],
\qquad\qquad\qquad
R(z) \simeq 3 M H'(z) \simeq \dfrac{2M^2}{z} \cos (2z).
\end{equation}

\noindent This is the path to show that $R''$ term dominates \cite{reheating}, so the equation of $X_k$ can be written as

\begin{equation}
X_k'' + \left[ \left(\dfrac{2k}{aM}\right)^2 + \left(\dfrac{2m}{M}\right)^2 - \xi \dfrac{R''}{M^2}\right] X_k = 0,
\end{equation}

\noindent which may be written in the form of the Mathieu equation

\begin{equation}
X_k'' + \left[ A_k - 2 q \cos(2z) \right] X_k = 0, \qquad\qquad
A_k = \left(\dfrac{2k}{aM}\right)^2 + \left(\dfrac{2m}{M}\right)^2,
\qquad\qquad q = \dfrac{4\xi}{z}.
\end{equation}

The interesting fact is that with a \ac{NMC} term in the action, the resulting equation is the same as the one obtained in \eqref{eq:mathieu_eq_starobinsky}. The results are, therefore, identical to the ones mentioned in section \ref{sec:reheating}.

% %%%%%%%%%%%%%%%%%%%%%%%%%%%%%%%%%%%%%%%%%%%%%%%%%%%%%%%%%%%%%%%%%%%%%%

\section{Solar System Long Range Regime}
\label{sec:impact_nmc}

Moving forward the parallelism with the pure $f(R)$ case from chapter \ref{cap:fofR}, the \ac{NMC} impact at the Solar System scales is analysed. As in section \ref{sec:solar_sys_impact}, matter is assumed to behave as dust, described by Eq. \eqref{eq:conditions_energy_mom_chap2}, so that the trace of the energy-momentum tensor is $T = -\rho c^2$ and the Lagrangian density of matter is treated as $\LL_m = -\rho c^2$.

The function $\rho = \rho(r)$ is again that of a spherically symmetric body with a static radial mass density $\rho = \rho(r)$. This function $\rho(r)$ and its first derivative are assumed to obey Eqs. \eqref{conditions for fe_march}.

The metric used is once more the perturbed \ac{FLRW} metric, repeated here for convenience:

\begin{equation}
ds^2 = - \left[1 + 2\Psi (r) \right] dt^2 + a(t)^2 \left( \left[1 + 2 \Phi(r) \right] dr^2 + r^2 d\Omega^2 \right),
\label{eq:metric_march}
\end{equation}

\noindent where $\Psi$ and $\Phi$ are perturbing functions such that $\left|\Psi(r)\right| \ll 1$ and $\left|\Phi(r)\right| \ll 1$. The scalar curvature of this spacetime is also expressed as the sum of a time evolving cosmological contribution $R_0(t)$ and a perturbation $R_1(r)$, induced by the local matter distribution,

\begin{equation}
R(r,t) \equiv R_0 (t) + R_1 (r), ~~~~\qquad ~~~~ \left| R_1 (r) \right| \ll R_0 (t).
\label{eq:curvature_march}
\end{equation}

This assumption allows to expand the functions $f^i(R)$ around $R=R_0$ and neglect all the nonlinear terms in $R_1$. This approximation is the same as the one detailed in Eq. \eqref{eq:conditions_linearization_chiba}, with the exception that now both $f^1(R)$ and $f^2(R)$ are subject to it.

\subsection{Solution for the curvature \texorpdfstring{$R_1$}{R1}}

The trace of the field equations \eqref{eq:field_equations_nmc} is
\begin{equation} \label{eq:trace_of_field_equations_march}
\left(f^1_R + 2 f^2_R \LL_m \right) R - 2 f^1 =-3 \square
\left(f^1_R + 2 f^2_R \LL_m \right)
+ \left( 1+ f^2 \right) T.
\end{equation}

\noindent Considering the decomposition of the energy-momentum tensor as $T_{\mu\nu} = T^\text{cos}_{\mu\nu} + T^\text{s}_{\mu\nu}$ and of the matter density as $\rho = \rho^\text{cos} (t) + \rho^\text{s} (r)$, the trace equation for the background curvature has the following solution

\begin{equation}
\left(f^1_{R0} + 2 f^2_{R0} \LL^\text{cos}_m \right) R_0 - 2 f_0^1 + 3 \square \left( f^1_{R0} + 2 f^2_{R0} \LL^\text{cos}_m \right) = \left(1 + f^2_0 \right) T^\text{cos}.
\label{eq:solution_trace_eq_background_march}
\end{equation}

Inserting this equation into \eqref{eq:trace_of_field_equations_march} and taking into account that all terms of order $\OO(R_1^2)$ or greater are neglected, yields

\begin{equation}\label{eq:R1_equation_march}
\begin{split}
\left[ -f^1_{R0} +  f^2_{R0}\LL_m + \left( f^1_{RR0} + 2 f^2_{RR0}\LL_m \right) R_0 \right] R_1 + & 3 \square \left[\left( f^1_{RR0} + 2 f^2_{RR0}\LL_m \right) R_1 \right] = \\ 
& \left( 1 +  f^2_0 \right) T^{\rm s} - 2 f^2_{R0}\LL_m^{\rm s}R_0 -
6\square\left( f^2_{R0}\LL_m^{\rm s} \right).
\end{split}
\end{equation}

The expansion of the operator $\square$ and neglecting products of $R_1$ or of its spatial derivatives with $\Psi$ and $\Phi$ and their spatial derivatives, gives the curvature equation the expression

\begin{equation}
\label{eq:U-equation_march}
\nabla^2 U -m^2 U = -\frac{1}{3}\left( 1 +  f^2_0 \right) \rho^{\rm s} +
\frac{2}{3} f^2_{R0}\rho^{\rm s}R_0 + 2 \rho^{\rm s} \square f^2_{R0} +
2 f^2_{R0} \nabla^2 \rho^{\rm s},
\end{equation}

\noindent where the potential $U(r,t)$ is defined as

\begin{equation}\label{eq:potentialdef}
U(r,t) = \left[ f^1_{RR0}(t) + 2 f^2_{RR0}(t)\LL_m(r,t)\right] R_1(r),
\end{equation}
and the mass parameter is
\begin{equation}
\label{eq:mass-formula_march}
m^2 =  \frac{1}{3}\left[\frac{f^1_{R0} -  f^2_{R0}\LL_m}{f^1_{RR0} + 2 f^2_{RR0}\LL_m}
- R_0 - \frac{3\square\left( f^1_{RR0} - 2 f^2_{RR0}\rho^{\rm cos}\right)
- 6\rho^{\rm s}\square f^2_{RR0}}{f^1_{RR0} + 2 f^2_{RR0}\LL_m} \right],
\end{equation}

\noindent where $f^1_{RR0} + 2 f^2_{RR0}\LL_m \neq 0$ is assumed. Naturally, this expression becomes the one obtained for the pure $f(R)$ theories, Eq. \eqref{eq:trace_equation_chiba_final}, when $f^2(R) = 0$. 

As a first step to work out the solution of the above equation, it is better to consider it outside the star, where $\rho^\text{s} = 0$, so that the mass parameter depends only on time, $m=m(t)$. Equation \eqref{eq:U-equation_march} thus becomes

\begin{equation}
\nabla^2 U = m^2(t) U,
\end{equation}

\noindent and its solution is a Yukawa potential of the type

\begin{equation}
U \sim \dfrac{e^{-mr}}{r},
\end{equation}
or, if $m^2 < 0$, an oscillating potential with strength $\sim 1/r$. 

The condition that $m^2 $ is positive-defined expands upon the previous stability constraint $f_{RR}^1 + f_{RR}^2 \LL_m > 0 $ \cite{Faraoni,Sequeira}, which is required to avoid the related Dolgov-Kawasaki instabilities \cite{DK}, and can be interpreted physically as requiring that gravity is attractive; the strong, weak, null and dominant energy conditions were also addressed in Ref. \cite{Sequeira}, showing that a wide class of NMC models is allowed; in fact, a NMC allows for a minimisation of the known violation of the null energy condition in wormholes \cite{LoboGarcia}.

The long range regime condition $mr \ll 1$ means that $U \sim 1 / r$ and so it is possible to neglect the term $m^2 U$ in equation \eqref{eq:U-equation_march}, when computing the total solution of $U(r,t)$. It follows then that equation \eqref{eq:U-equation_march} becomes 

\begin{equation}\label{U-equation-simplified}
\nabla^2 U = \eta(t)\rho^{\rm s}(r) + 2 f^2_{R0} \nabla^2 \rho^{\rm s},
\end{equation}
with
\begin{equation}\label{etadef}
\eta(t) = -\frac{1}{3}\left( 1 + f^2_0 \right) + \frac{2}{3} f^2_{R0}R_0 + 2 \square f^2_{R0}.
\end{equation}

Computing the solution of $U(r,t)$ outside the body, $r > R_S$, one obtains

\begin{equation}\label{eq:U-equation-simplified-outside}
U(r,t) = -\frac{\eta(t) }{ 4\pi} \frac{M_{\rm S}}{ r},
\end{equation}

\noindent where $M_S$ is the total mass of the spherical body. Through Eq. \eqref{eq:potentialdef}, the curvature $R_1$ is defined as 
\begin{equation}\label{eq:R1-solution_march}
R_1(r,t) = \frac{\eta(t) }{ 4\pi \left( 2 f^2_{RR0}\rho^\text{cos}-  f^1_{RR0}\right)} \frac{M_\text{S}}{r},
\end{equation}

\noindent which obviously reduces to Eq. \eqref{eq:curvature_solution_chiba}   when $f^2(R) = 0$. Also, notice that the variation of $R_1$ in time is on a time scale much bigger than the one proper from the Solar System, so that it is plausible to assume $R_1(r,t) \simeq R_1(r)$. The $R_1(r)$ solution for $r < R_S$ is not relevant for the following analysis, so it is not presented here (for details, check Ref. \cite{solar}).

\subsection{Solution for \texorpdfstring{$\Psi$}{Psi} and \texorpdfstring{$\Phi$}{Phi}}

The same linearization procedure that was used for the curvature may be used for the $tt$ and $rr$ components of the field equations \eqref{eq:field_equations_nmc}. The computation of $\Psi$ once more goes through a decomposition of the type $\Psi = \Psi_0 + \Psi_1$ and the computation of $\Phi$ is nothing more than a simple integration. The development of these computations is beyond the scope of this section analysis, whose main point is simply to present the results of the long range Solar System effect of the \ac{NMC}. Nonetheless, for more details one should resort to Ref. \cite{solar}.

The final solution for $\Psi$ outside the spherical body is,
\begin{equation}\label{Psi-solution}
\Psi(r,t) = - \frac{ 1 +  f^2_0 +  f^2_{R0}R_0 + 3\square f^2_{R0} }
{6\pi \left( f^1_{R0} - 2 f^2_{R0}\rho^{\rm cos} \right)} \,\frac{M_{S}}{r},
\qquad r \geq R_S,
\end{equation}
which, if $f^2(R)=0$, is reduced to the solution of $\Psi$ from section \ref{subsec:long-range}. This expression can be said to include a gravitational coupling slowly varying in time,
\begin{eqnarray}\label{G-constant}
G_\text{eff} &=& \frac{ \omega(t) }
{6\pi \left( f^1_{R0} - 2 f^2_{R0}\rho^{\rm cos} \right)}, \\ \nonumber
\omega(t) &=& 1 +  f^2_0 +  f^2_{R0}R_0 +3\square f^2_{R0}.
\end{eqnarray}
This expression shows that the timescale $\dot{G}_\text{eff}/G_\text{eff}$ is much bigger than the timescale of the Solar System, and thus the approximations $G_\text{eff} \simeq {\rm const.}$ and $\Psi(r,t) \simeq \Psi(r)$ can be made. The available bounds on $\dot{G}_\text{eff}/G_\text{eff}$  \cite{ChibaReview} may help to constrain $f^1(R)$ and $f^2(R)$ through Eq. \eqref{G-constant}. The Newtonian limit also requires that $\Psi(r)$ should be proportional to $M_S \slash r$, leading to the following constraint on the functions $f^1(R)$ and $f^2(R)$:

\begin{equation}\label{Newt-limit}
\left\vert 2 f^2_{R0} \right\vert \rho^\text{s}(r) \ll
\left\vert f^1_{R0} - 2 f^2_{R0}\rho^\text{cos}(t) \right\vert, \qquad r \leq R_S.
\end{equation}

On the other hand, together with the obtained solutions of $R_1$ and $\Psi$, the $rr$ component of the field equations yields for $\Phi$ the solution

\begin{equation}
\Phi(r,t) = \frac{1 +  f^2_0 + 4 f^2_{R0}R_0 + 12\square f^2_{R0}}
{12\pi \left( f^1_{R0} - 2 f^2_{R0}\rho^{\rm cos} \right)} \,\frac{M_{\rm S}}{r},
\end{equation}

\noindent where $\Phi(r,t) \simeq \Phi(r)$ is once more valid. When $f^2(R)=0$, the solution for $\Phi$ becomes the one from the pure $f(R)$ gravity, \eqref{eq:phi_eq_chiba}.

Having the solutions of both potentials $\Psi$ and $\Phi$, the PPN parameter $\gamma$ is then
\begin{equation}\label{gamma}
\gamma = \frac{1}{2} \, \left[\frac{1 +  f^2_0 + 4 f^2_{R0}R_0 + 12\square f^2_{R0}}
{1 +  f^2_0 +  f^2_{R0}R_0 +3\square f^2_{R0}} \right],
\end{equation}
showing that it is completely defined by the background metric $R_0$, whose value is obtained from the cosmological solution of \ac{NMC} gravity, developed in Ref. \cite{nmcexpand}. The case $f^2(R)=0$ obviously yields the value $\gamma = 1\slash 2$ from the computations of section \ref{subsec:long-range}. 
However, it should be said that this expression for $\gamma$ does not work properly when the functions $f^i (R)$ reduce to their GR expressions, since it implies that the mass parameter $m$, defined in (\ref{eq:mass-formula_march}), is ill-defined (and divergent), so that many assumptions made to compute $\Psi$ and $\Phi$ are not satisfied \cite{solar}.

Concluding the chapter, all of what was said above means that for a \ac{NMC} model in a long range regime to be compatible with Solar System results, one of these conditions has to be satisfied:
\begin{itemize}
\item[{\rm (i)}] The present analysis is not adequate to test the validity of the model, because either the condition $|mr| \ll 1$ at Solar System scales shouldn't apply or the nonlinear terms in $R_1$ should be considered in the Taylor expansions from (\ref{eq:conditions_linearization_chiba}) ($i.e.$ the condition $\left\vert R_1 \right\vert \ll R_0$ is not satisfied);
\item[{\rm (ii)}] Or, if both conditions from point (i) are otherwise satisfied, then the validity condition of the Newtonian limit (\ref{Newt-limit}) has to be satisfied, so that the value of $\gamma$ (\ref{gamma}) needs to satisfy the Cassini measurement constraint $\gamma = 1 + (2.1 \pm 2.3) \times 10^{-5}$ \cite{cassini}.
\end{itemize}

To advance with condition (ii), the parameter of mass $m^2$ (\ref{eq:mass-formula_march}) should be computed with the cosmological solution, \textit{i.e.} with the values of $R_0(t)$ and $\rho^{\rm cos}(t)$. For that, the functions $f^i(R)$ can in general assume power-law forms compatible with the description of the current phase of accelerated expansion of the Universe \cite{nmcexpand,capozziello3}, such as

\begin{equation}
f^1(R) = 2\kappa R, ~~~~~\qquad~~~~~ f^2(R) = \left(\dfrac{R}{R_n}\right)^{-n}, ~~~~ n>0.
\end{equation}

The development of the conditions presented in point (i) in this power-law form of \ac{NMC} gravity is detailed in Ref. \cite{solar}, where the authors make the following conclusions:

\begin{itemize}
\item The Solar System long-range regime $|m| r \ll 1$ leads to $n \gg 10^{-25}$;
\item The validity condition for the Newtonian limit \eqref{Newt-limit} implies that $n \ll 10^{-33}$;
\item The perturbative regime $\left\vert R_1 \right\vert \ll R_0$ leads to an unphysical fine tuning of the density profile $\rho^\text{s}(r)$ inside the spherical body.
\end{itemize}

\noindent The last conclusion is related to the lack of validity of the perturbative regime and it applies to a wide range of astrophysical objects, as long as they are spherical bodies with whom the weak-field approximation can be used. This conclusion means that the procedure used, first applied in section \ref{subsec:long-range}, does not exclude or even constrain the \ac{NMC} model. It is a null-result that nevertheless prompts to inspect more closely the inverse situation of a relevant mass parameter and a negligible cosmological background.

\cleardoublepage

% %%%%%%%%%%%%%%%%%%%%%%%%%%%%%%%%%%%%%%%%%%%%%%%%%%%%%%%%%%%%%%%%%%%%%%
% Dummy Chapter:
% %%%%%%%%%%%%%%%%%%%%%%%%%%%%%%%%%%%%%%%%%%%%%%%%%%%%%%%%%%%%%%%%%%%%%%

% %%%%%%%%%%%%%%%%%%%%%%%%%%%%%%%%%%%%%%%%%%%%%%%%%%%%%%%%%%%%%%%%%%%%%%
% Gravitational constraints:
% %%%%%%%%%%%%%%%%%%%%%%%%%%%%%%%%%%%%%%%%%%%%%%%%%%%%%%%%%%%%%%%%%%%%%%
\chapter{Experimental Constraints on Modifications of \ac{GR}}
\label{cap: gravity constraints}

\section{Tests of the Gravitational \ac{ISL}}
\label{sec:isl}

The law of gravity developed by Sir Isaac Newton in the beginning of the 18$^\text{th}$ century and generalized in Einstein's \ac{GR} stands as the first fundamental law of physics to be discovered (followed by electromagnetism and the weak and strong forces of quantum mechanics). Its consistency with the phenomena of nature is amazingly strong, from the small scale of a few milimeters all through to the cosmological scale. The standard gravitational \ac{ISL}, originally formalized by Newton, is

\begin{equation}
F(r) = \dfrac{G m_i m_j}{r^2},
\end{equation}

\noindent where $G$ is Newton's constant of gravitation and $r$ is the distance between the point masses $m_i$ and $m_j$. This expression also stems from the \ac{GR} theory, in the nonrelativistic limit of weak fields.

There are, of course, some problems with the accepted \ac{GR} theory of gravity, already identified in section \ref{sec:mod_problems}. These are related to the need for dark matter, the present accelerated expansion of the Universe and to the compatibility of \ac{GR} with quantum mechanics. Moreover, there are some phenomena arising from the solutions for this latter problem, which may be detected at the scale level of Newton's gravitational law. These phenomena, mostly enumerated in Refs. \cite{fischbach, adelberg1}, are often related to forces not described in an \ac{ISL} that should also account for the effects of gravity. Indeed, it is now usual to identify and constrain these deviations of the \ac{ISL} through a Yukawa contribution to the potential,

\begin{equation}
U(r) = -G \dfrac{m_i}{r} \left(1 + \alpha e^{-r/\lambda} \right),
\end{equation}

\noindent where $\alpha$ is called the Yukawa strength parameter and $\lambda$ is a characteristic length scale. The motivation for this Yukawa potential comes from the presence of a massive scalar field, so that it is simply the solution of Klein-Gordon's equation $\square \phi = m^2 \phi$ \cite{adelberg1}, with $\lambda=1/m$. It should also be noticed that when $\lambda \to \infty$, one obtains a modification on the Newton constant $G_N = G \left( 1 + \alpha \right)$.

\begin{figure}[ht]
\centering
\includegraphics[width=\textwidth]{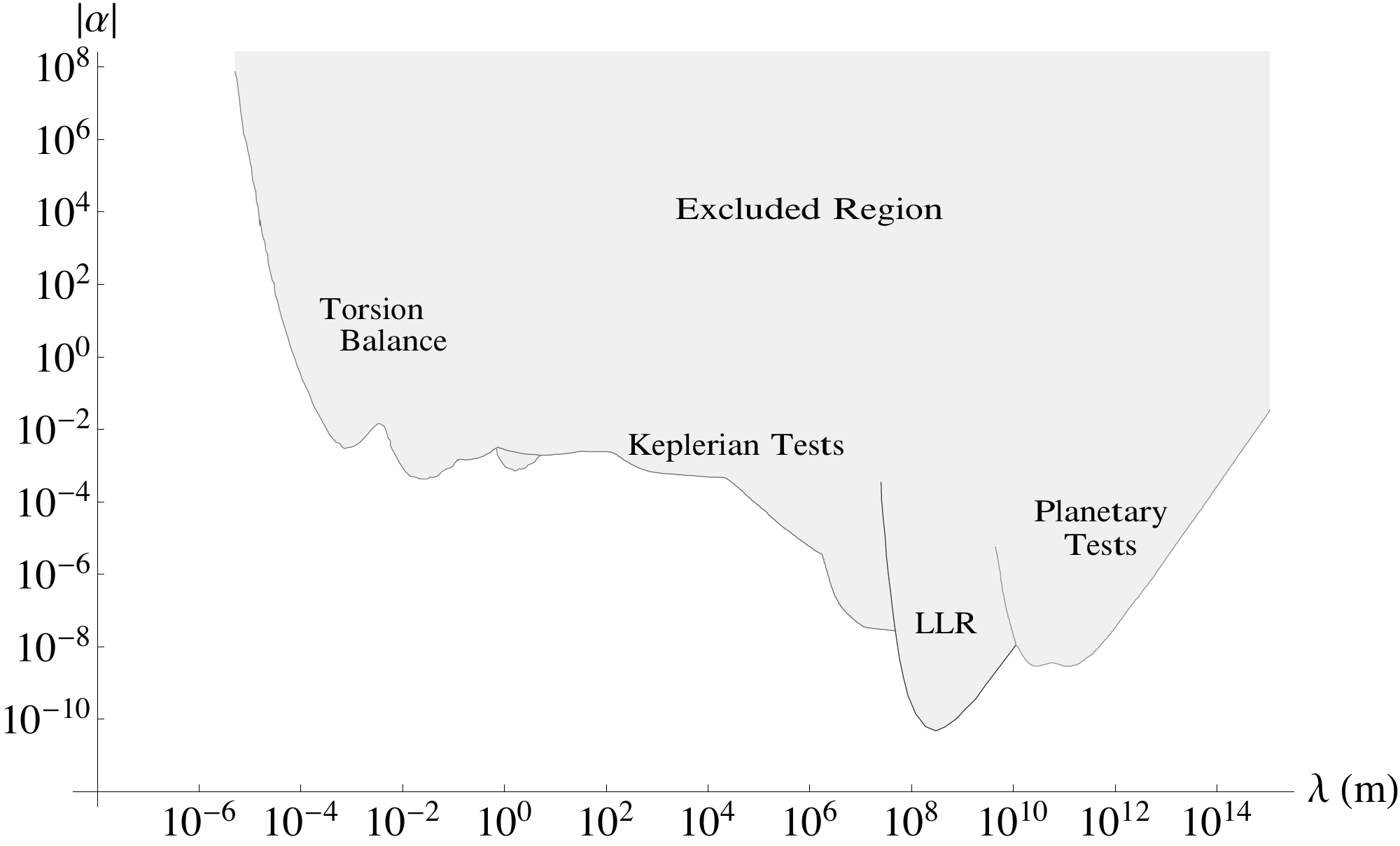}
\caption{Yukawa exclusion plot for $\alpha$ and $\lambda$. Adapted froms Refs. \cite{adelberg1,salumbides}.}
\label{fig:yukawa_plot}
\end{figure}

These parameters $\alpha$ and $\lambda$ are constrained by limits imposed from experimental results and figure \ref{fig:yukawa_plot} shows how large $\alpha$ can be in the different length scales dictated by $\lambda$. The overall data in this graphic is a composition from different experiments. The values of $\alpha$ above the curve, in the shaded area, are excluded as physically possible values with a $95 \%$-confidence level.

The results for the left curve are determined by the torsion balance experiments from the E\"ot-Wash group, presented in Ref. \cite{adelberg3} and Refs. therein. The Keplerian and \ac{LLR} tests represent the laser ranging experiments between the Earth, the Moon and between each of these and the LAGEOS satellite, whose results can be found in Ref. \cite{fischbach}. The planetary results do not differ ever since the ones obtained by the analysis of varied astronomical data in 1988 were made, which are more detailed in Ref. \cite{talmadge}.

% %%%%%%%%%%%%%%%%%%%%%%%%%%%%%%%%%%%%%%%%%%%%%%%%%%%%%%%%%%%%%%%%%%%%%%

\section{\ac{PPN} Formalism}
\label{sec:ppn}

Most of modern theories of gravity work only with two different elements: the matter and the metric. The action of gravity in the Universe is then a direct consequence of the metric guiding the motion of matter, which acts as source to the latter. This is important, for instance, to compare competing theories, which become particularly simple when the studies are made in the slow motion and weak field limit - an approximation known as the post-Newtonian limit. This name is very appropriate because the post-Newtonian limit is simply a way to write a general theory of gravitation in the form of lowest-order deviations from the Newton law of gravitation.

Mathematically, this limit can be written as an expansion of the Minkowski metric in terms of small gravitational potentials, that are defined in terms of matter variables like the matter density $\rho({\bf x})$, just like the Newtonian gravity potential, where $\mathbf{x}$ is the position vector. Through the years this expansion has taken several forms, initially developed by Nordtvedt \cite{nordtvedt} and then generalized by Will in the standard form known today as the \ac{PPN} formalism \cite{willbook}.

Accordingly, the current version of the \ac{PPN} formalism is written with a metric of the form

\begin{equation}
\begin{split}
g_{00} = &-1 + 2U - 2\beta U^2 -2\xi \Phi_W + \left( 2\gamma +2 + \alpha_3 + \zeta_1 -2\xi \right) \Phi_1 + 2 \left(3\gamma -2\beta +1 + \zeta_2 + \xi \right) \Phi_2\\
& + 2 \left(1+\zeta_3\right) \Phi_3 + 2\left(3 \gamma + 3\zeta_4 -2\xi\right) \Phi_4 - \left(\zeta_1 - 2 \xi \right) \mathcal{A} - \left(\alpha_1-\alpha_2-\alpha_3 \right) w^2 U - \alpha_2 w^i w^j U_{ij}\\
& + \left(2 \alpha_3 -\alpha_1 \right)w^i V_i + \OO \left(\epsilon^3 \right), \\
g_{0i} = &-\dfrac{1}{2} \left(4\gamma +3 +\alpha_1 - \alpha_2 + \zeta_1 - 2\xi \right) V_i - \dfrac{1}{2} \left(1 + \alpha_2 - \zeta_1 + 2\xi \right) W_i - \dfrac{1}{2}\left(\alpha_1 - 2 \alpha_2 \right) w^i U \\
&-\alpha_2 w^j U_{ij} + \OO \left( \epsilon^{5/2} \right), \\
g_{ij} =& \left(1 + 2 \gamma U \right) \delta_{ij} + \OO\left(\epsilon^2\right),
\end{split}
\label{eq:ppn_formalism}
\end{equation}

\noindent where $i,j = \{1,2,3\}$, the parameter $\epsilon$ determines the order of smallness, with $U \sim v^2 \sim p/\rho \sim \epsilon$ and $| d/dt| / |d/dx| \sim \epsilon^{1/2}$ and with relativistic units $G=c=1$, $U$ is a gravitational potential, $v$ the matter velocity, $\rho$ the matter density and $p$ the pressure of the matter in its comoving frame.

The ten \ac{PPN} parameters $\gamma, \beta, \xi, \alpha_1, \alpha_2, \alpha_3, \zeta_1, \zeta_2, \zeta_3, \zeta_4$ represent the coefficients of the metric potentials $U, U_{ij}, \Phi_W, \mathcal{A}, \Phi_1, \Phi_2, \Phi_3, \Phi_4, V_i, W_i$ and may take different values according to the theory being studied. Each of the parameters measures or indicates general properties of the specific theories of gravity. For instance, $\gamma$ measures the space-curvature produced by a unit rest mass, $\beta$ measures the nonlinearity in the superposition law for gravity, the $\zeta_i$ parameters relate to the violation of conservation of the total momentum and the others yield specific effects related to the existence of preferred frames and Lorentz violation. Interestingly enough, the \ac{GR} theory only includes the parameters $\gamma=\beta=1$, all the others are null, therefore the metric \ac{GR} in a resting frame takes the form

\begin{equation}
\begin{split}
g_{00} =& -1 + 2U - 2 U^2 + \OO\left(\epsilon^3\right), \\
g_{0i} =& 0, \\
g_{ij} =& \left( 1 +2 U \right) \delta_{ij} + \OO\left(\epsilon^2\right),
\end{split}
\end{equation}

\noindent where the cross components $g_{0i}$ vanish because the frame is at rest (the potential $V_i = 0$, for it depends on the velocity of matter) and $U$ is defined as the standard Newtonian gravity potential

\begin{equation}
U \left( r,t \right) \equiv \int \dfrac{\rho\left(r',t'\right)}{\left| r- r' \right|} d^3x'.
\end{equation}

The experimental data of the Solar System validates the \ac{GR} theory, so a different theory to be physically valid needs to predict values for the \ac{PPN} parameters very close to those given by \ac{GR}, i.e. $\beta \sim \gamma \sim 1$ and all others very close to zero. More specifically, if one wants to check the validity of a theory, the best way to start would be to compute the value of $\gamma$, determined by the Cassini experiment as $\gamma = 1 + (2.1 \pm 2.3) \times 10^{-5}$ \cite{cassini}. A simple way to compute the \ac{PPN} parameter $\gamma$ is to derive the metric elements

\begin{equation}
g_{00} = -\left(1 + 2\Psi \right), \qquad\qquad g_{11} = 1 + 2\Phi,
\end{equation}

\noindent where $\Psi$ and $\Phi$ are small perturbations of the form of a Newtonian potential; by the \ac{PPN} formalism \eqref{eq:ppn_formalism}, $\Psi \to -U$ and $\Phi \to \gamma U$, therefore the parameter $\gamma$ can be determined by

\begin{equation}
\gamma = - \dfrac{\Phi}{\Psi}.
\end{equation}

\noindent For example, \ac{BD} theories predict that

\begin{equation}
\gamma = \dfrac{1+\omega}{2+\omega},
\end{equation}

\noindent so that one must have $\omega \gtrsim 4 \times 10^5$.

% %%%%%%%%%%%%%%%%%%%%%%%%%%%%%%%%%%%%%%%%%%%%%%%%%%%%%%%%%%%%%%%%%%%%%%

\section{Geodetic Precession}
\label{sec:geodetic}

In classical mechanics, the rotation of a rigid body is described by the torque equation

\begin{equation}
\boldsymbol{ \tau } = \dfrac{d \mathbf{L} }{dt}, \qquad\qquad \text{with} \qquad \mathbf{L} = I \boldsymbol{\omega},
\end{equation}

\noindent where $\boldsymbol{\tau}$ is the torque vector\footnote{In this section, the vectors are identified with the respective variables written in bold such as $\vec{r}\equiv \mathbf{r}$.}, $\mathbf{L}$ is the angular momentum vector, $I$ is the moment of inertia and $\boldsymbol{\omega}$ is the angular velocity, defined as $\boldsymbol{\omega}= d\boldsymbol{\theta} / dt$. From these equations, it is possible to write the conservation of angular momentum, $i.e.$ the case where no rotational forces are applied to the body, yielding a constant value for it,

\begin{equation}
\dfrac{d\mathbf{L}}{dt} = 0.
\end{equation}

\begin{figure}[ht]
\centering
\includegraphics[width=0.5\textwidth]{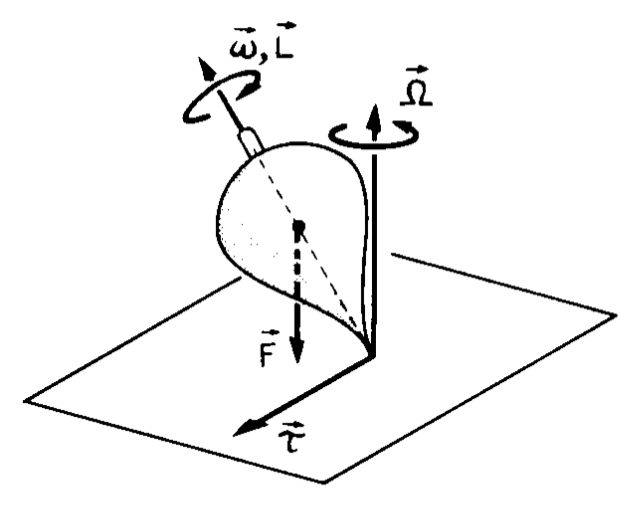}
\caption{A rapidly spinning top \cite{feynman}.}
\label{fig:spinningtop}
\end{figure}

The phenomenon of precession according to these equations, commonly seen in gyroscopes and spinning tops, happens when a torque is applied to a rotating body with constant angular momentum $\mathbf{L}_0$. The spinning top case, from figure \ref{fig:spinningtop}, is a good example to understand the mathematics of precession. 

The spinning top remains with the angular momentum conserved, $d\mathbf{L}_0/dt=0$, and rotation speed $\boldsymbol{\omega}$ and no torque is acted on it until it reaches the ground. When it lands, the back reaction to the weight $\mathbf{F}$ yields a torque $\boldsymbol{\tau}$ that breaks the conservation of the angular momentum. The new equation for the movement of the top is then

\begin{equation}
\boldsymbol{\tau} = \dfrac{d\mathbf{L}}{dt} = \boldsymbol{\Omega} \times \mathbf{L_0},
\end{equation}

\noindent where $\boldsymbol{\Omega}$ is a rotation velocity different from $\boldsymbol{\omega}$, which characterizes the precession movement of the spinning top.

These computations can be generalized to a curved space-time by the principle of General Covariance, with $d/dx^\mu \to \nabla_\mu$. Nonetheless, \ac{GR} predicts a different kind of precession for freely falling spinning bodies, $i.e.$ bodies where no torque or any other force is being applied. Following Ref. \cite{weinberg}, consider a gyroscope with an intrinsic angular momentum vector $S^\mu = \left( S^0, \mathbf{S}\right)$ orbiting a spherical body like the Earth. The equation of parallel transport for $S^\mu$ becomes

\begin{equation}
\nabla_\nu S_\mu = 0 \Leftrightarrow \dfrac{d S_\mu}{d\tau} = \Gamma^\lambda_{\mu\nu} S_\lambda \dfrac{dx^\nu}{d\tau},
\label{eq:paralleltransport}
\end{equation}

\noindent where $\tau$ is the coordinate of proper time. The angular momentum vector $S_\mu$ is defined so as to be orthogonal to the velocity vector,

\begin{equation}
\dfrac{dx^\mu}{d\tau} S_\mu = 0 \Rightarrow S_0 = -v^i S_i,
\end{equation}

\noindent where $i \in \{1,2,3\}$. Inserting this condition into \eqref{eq:paralleltransport} and multiplying it with $d\tau / dt$ yields

\begin{equation}
\dfrac{dS_i}{dt} = \Gamma^j_{i0} S_j - \Gamma^0_{i0} v^j S_j - \Gamma^j_{ik} v^k S_j - \Gamma^0_{ik} v^k v^j S_j
\simeq \left( ^{(3)}\Gamma^j_{i0} - ^{(2)}\Gamma^0_{i0} v^j + ^{(2)}\Gamma^j_{ik} v^k \right) S_j.
\label{eq:parallel_approx}
\end{equation}

\noindent The computation of this equation was done with a metric according to the \ac{PPN} formalism\footnote{In this subsection all the computations are made considering $c=1$.},

\begin{equation}
g^{00} = -1 -2 \phi, \qquad\qquad g^{i0} = \zeta_i, \qquad\qquad g^{ij} = \left(1-2\phi\right) \delta_{ij}.
\end{equation}

\noindent The approximation from Eq. \eqref{eq:parallel_approx} considered only parameters of the order $\phi \sim \OO(\epsilon)$ and $\zeta_i \sim \OO(\epsilon^{3/2})$ or less, where the parameter $\epsilon$ was defined after Eq. \eqref{eq:ppn_formalism}. The number preceding some terms like $^{(n)}\Gamma$ represents the order of the term,  $\Gamma = \OO(\epsilon^n)$. Expanding the above equation with the connection terms from the metric yields

\begin{equation}
\dfrac{d\mathbf{S}}{dt} \simeq \dfrac{1}{2} \mathbf{S} \times \left( \nabla \times \boldsymbol{\zeta} \right) - \mathbf{S}\dfrac{\partial \phi}{\partial t} - 2 \left(\mathbf{v} \cdot \mathbf{S}\right) \nabla \phi - \mathbf{S} \left( \mathbf{v} \cdot \nabla \phi \right) + \mathbf{v} \left( \mathbf{S} \cdot \nabla \phi \right).
\label{eq:rate_of_change_spin_S}
\end{equation}

It is useful at this stage to introduce a new spin vector $\boldsymbol{\mathcal{S}}$ defined as 

\begin{equation}
\mathbf{S} = \left( 1- \phi \right) \boldsymbol{\mathcal{S}} + \dfrac{1}{2} \mathbf{v} \left( \mathbf{v} \cdot \boldsymbol{\mathcal{S}} \right) \quad \Leftrightarrow \quad
\boldsymbol{\mathcal{S}} = \left( 1+ \phi \right) \mathbf{S} - \dfrac{1}{2}  \mathbf{v} \left( \mathbf{v} \cdot \boldsymbol{S} \right),
\end{equation}

\noindent where its equation of parallel transport to order $\OO(\epsilon^{3/2})$ or less is

\begin{equation}
\dfrac{d\boldsymbol{\mathcal{S}}}{dt} = \dfrac{d\mathbf{S}}{dt} + \mathbf{S}\left( \dfrac{\partial \phi}{\partial t} + \mathbf{v} \cdot \nabla \phi \right) + \dfrac{1}{2} \nabla \phi \left( \mathbf{v} \cdot \mathbf{S}\right) + \dfrac{1}{2} \mathbf{v} \left( \mathbf{S} \cdot \nabla \phi \right).
\end{equation}

\noindent $\mathbf{S}$ was considered as constant everywhere in space and $d\mathbf{v}/dt \simeq - \nabla \phi$. Inserting Eq. \eqref{eq:rate_of_change_spin_S} into this one yields for $\boldsymbol{\mathcal{S}}$

\begin{equation}
\dfrac{d \boldsymbol{\mathcal{S}}}{dt} = \boldsymbol{\Omega} \times \boldsymbol{\mathcal{S}}, \qquad\qquad
\boldsymbol{\Omega} = \dfrac{1}{2} \nabla \times \boldsymbol{\zeta} - \dfrac{3}{2}\mathbf{v} \times \nabla \phi,
\end{equation}

\noindent showing that $\boldsymbol{\mathcal{S}}$ precesses with angular velocity $\boldsymbol{\Omega}$.

As shown in Ref. \cite{weinberg}, if the central spherical body is the rotating Earth at rest, then the fields $\phi$ and $\boldsymbol{\zeta}$ can be written as

\begin{equation}
\phi = - \dfrac{G M_\oplus}{r}, \qquad\qquad \boldsymbol{\zeta} = \dfrac{2 G}{r^3} \left( \mathbf{r} \times \mathbf{J}_\oplus \right),
\end{equation}

\noindent where $M_\oplus$ and $\mathbf{J}_\oplus$ are the mass and angular momentum of the Earth. Inserting this fields into the precession velocity of $\boldsymbol{\mathcal{S}}$ means that it becomes

\begin{equation}
\boldsymbol{\Omega} = 3 G \dfrac{\mathbf{r} \left( \mathbf{r} \cdot \mathbf{J}_\oplus \right)}{r^5} - G \dfrac{\mathbf{J}_\oplus}{r^3} + 3 G M_\oplus \dfrac{\mathbf{r}\times \mathbf{v}}{2 r^3},
\end{equation}

\noindent where the last term depends only on the mass of the earth and not on its spin and it is called geodetic precession.

\begin{figure}[ht]
\centering
\includegraphics[width=\textwidth]{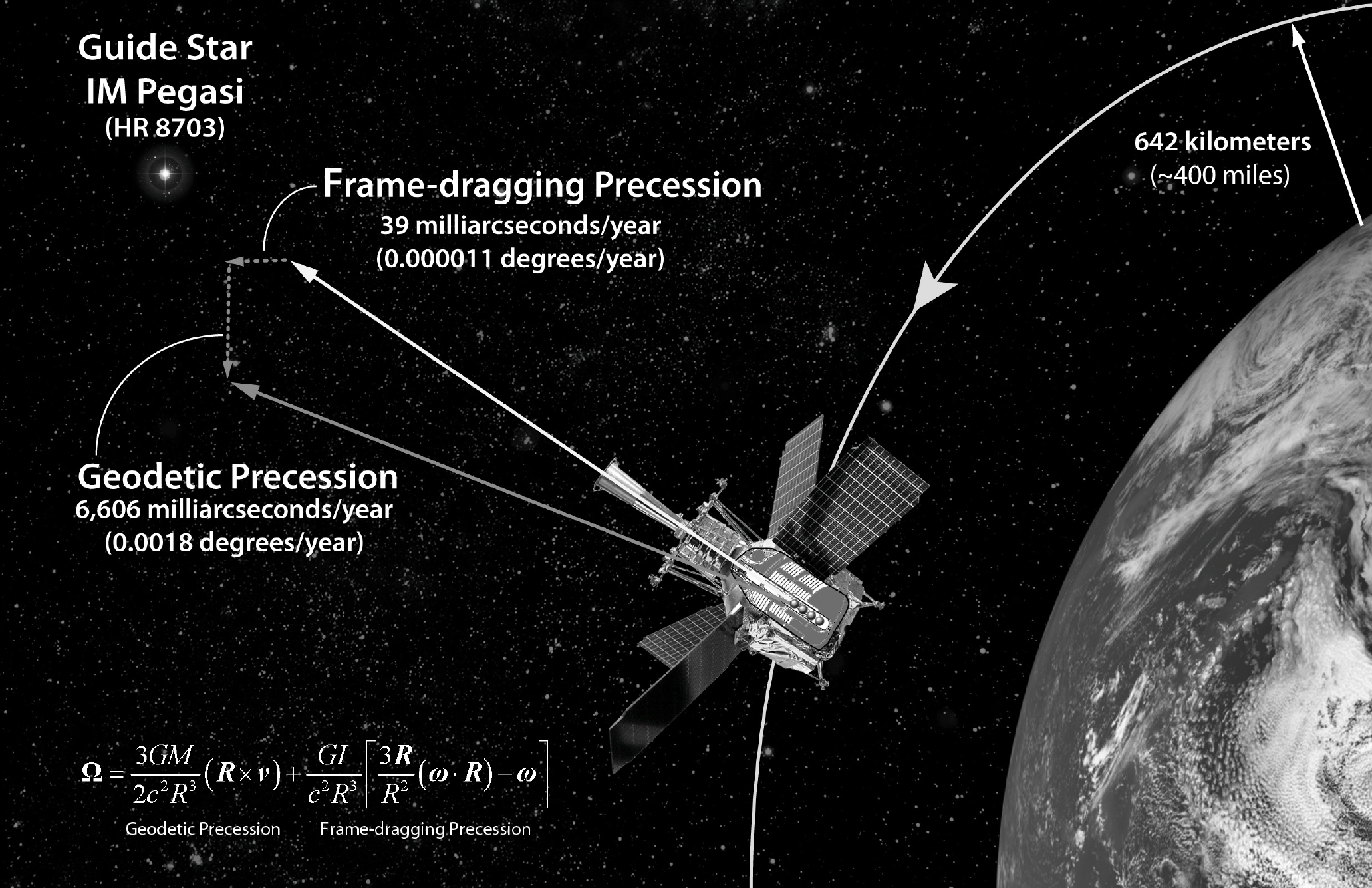}
\caption{Representation of the \ac{GPB} experiment and the geodetic effect prediction \cite{gpbteam}.}
\label{fig:gpb}
\end{figure}

The geodetic precession effect is a direct consequence of the \ac{GR} theory in the orbit of the Earth and it was observationally checked by the \ac{GPB} experiment. This experiment consisted in a satellite orbiting the Earth containing a telescope focused to the binary star HR8703, from the Pegasus constellation, to use as a position reference, and four extremely precise gyroscopes (four in order to decrease the data error). The spin axis of the gyroscopes is set to be aligned with the guide star during a certain period of time. At the end of it, the change in the precession of the spin axis alignment reveals the predicted geodetic effect. The display of the experiment is shown in figure \ref{fig:gpb}, which also shows another effect predicted by \ac{GR} that was tested in the same experiment $-$ the frame-dragging effect.

The \ac{GPB} confirmed the geodetic effect as a correct prediction of \ac{GR} with $0,28 \%$ of accuracy \cite{GPB}. Interactions between random patches of electrostatic potential in the gyroscopes did not allow for such a high precision in the frame-dragging effect \cite{gpbwill}.

Although not as accurate as the \ac{GPB} experiment, tests on \ac{LLR} also confirm the geodetic precession, with a 1\% accuracy \cite{LLR}. \ac{LLR} involves the analysis of data coming from lasers emitted from the Earth to the Moon and then reflected back to the Earth. This \ac{LLR} experiment also yields a specific constraint on Yukawa modifications to the \ac{ISL}, equivalent to the perigee precession of LAGEOS II \cite{lageos}, which will be mentioned in a later section. 

\cleardoublepage

% %%%%%%%%%%%%%%%%%%%%%%%%%%%%%%%%%%%%%%%%%%%%%%%%%%%%%%%%%%%%%%%%%%%%%%
% Dummy Chapter:
% %%%%%%%%%%%%%%%%%%%%%%%%%%%%%%%%%%%%%%%%%%%%%%%%%%%%%%%%%%%%%%%%%%%%%%

% %%%%%%%%%%%%%%%%%%%%%%%%%%%%%%%%%%%%%%%%%%%%%%%%%%%%%%%%%%%%%%%%%%%%%%
% f(R) theories:
% %%%%%%%%%%%%%%%%%%%%%%%%%%%%%%%%%%%%%%%%%%%%%%%%%%%%%%%%%%%%%%%%%%%%%%
\chapter{Short Range Solar System impact of a \ac{NMC}}
\label{cap:paper}

The contents of the following chapters include original work and are completely based in Ref. \cite{nunocb}.

In this chapter the procedure from section \ref{subsec:short-range} to assess the short range Solar System impact of $f(R)$ theories will be followed, where a \ac{NMC} model will be considered instead. This is done considering that the additional degree of freedom arising from a non-trivial $f(R)$ function is sufficiently massive so that its effects are not extremely long-ranged, and as such the background cosmological setting can be neglected $-$ this point (which shall be developed in the following sections) shows that the method is complementary to the recent study on the compatibility between cosmological and Solar System dynamics exposed in section \ref{sec:impact_nmc}.

The action functional used is again the one from Eq. \eqref{eq:action_nmc}, with the field equations \eqref{eq:field_equations_nmc}. It is once more assumed that matter behaves as dust, with the energy momentum tensor described by Eq. \eqref{eq:conditions_energy_mom_chap2}, and that the mass density respects the conditions from Eq. \eqref{conditions for fe_march}.

As before, the metric used is one that describes the spacetime around a spherical star like the Sun and it is given by the following perturbation of the Minkowski metric, in spherical coordinates:
\begin{equation} \label{metric}
ds^2=-\left[1+2\Psi\left(r\right)\right]c^2dt^2+\left[1+2\Phi(r)\right] dr^2 + r^2 d\Omega^2,
\end{equation}
\noindent where $\Psi$ and $\Phi$ are perturbing functions such that $\left|\Psi(r)\right| \ll 1$ and $\left|\Phi(r)\right| \ll 1$. For the purpose of the present work the functions $\Psi$ and $\Phi$ will be computed at order $\OO(1/c^2)$.
The functions $f^i(R)$ are assumed to admit the following Taylor expansions around $R=0$, which coincide with the forms used in the \ac{NMC} preheating study from section \ref{sec:preheating}:
\begin{equation} \label{f(R) equations}
f^1(R)=2\kappa \left(R+\dfrac{R^2}{6m^2}\right) + \OO(R^3), \qquad\qquad
f^2(R)=2\xi \dfrac{R}{m^2} + \OO(R^2),
\end{equation}
\noindent where $m$ is a characteristic mass scale and $\xi$ a dimensionless parameter specific of the \ac{NMC}, indicating the relative strength of the latter with respect to the quadratic term in $f^1(R)$.

It should be noted that the Cosmological Constant is dropped, consistent with the assumption that the metric is asymptotically flat --- {\it i.e.} no cosmological background with a time-dependent, non-vanishing curvature $R_0 \neq 0$ is assumed, contrary to what was considered in section \ref{sec:impact_nmc}. 

\section{Solution of linearized modified field equations}
%%%%%%%%%%%%%%%%%%%%%%%%%%%%%%%%%%%%%%%%%%%%%%%%%%%%%%%%%%%%%%%%%%

%%%%%%%%%%%%%%%%%%%%%%%%%%%%%%%%%%%%%%%%%%%%%%%%%%%%%%%%%%%%%%%%%%
\subsection{Solution for the curvature \texorpdfstring{$R$}{R}}
%%%%%%%%%%%%%%%%%%%%%%%%%%%%%%%%%%%%%%%%%%%%%%%%%%%%%%%%%%%%%%%%%%

The trace of the field equations \eqref{eq:field_equations_nmc} is
\begin{equation} \label{trace of field equations}
\left(f^1_R + 2 f^2_R \LL_m \right) R - 2 f^1 =-3 \square
\left(f^1_R + 2 f^2_R \LL_m \right)
+ \left( 1+ f^2 \right) T.
\end{equation}
\noindent After expanding the trace with the respective expressions, the equation is linearized: this is done by neglecting terms of order $\OO( 1 \slash c^3)$ or smaller, which then yields the following equation,
\begin{equation} \label{curvature equation}
\nabla^2 R - m^2 R = - \dfrac{8\pi G}{c^2} m^2
\left[ \rho -6\left(\dfrac{2\xi}{m^2}\right) \nabla^2 \rho \right].
\end{equation}
\noindent The variable substitution $u = r R$, enables the more straightforward expression
\begin{equation}
\dfrac{d^2 u(r)}{dr^2} - m^2 u(r) = s(r),
\label{curvature equation, variable substitution}
\end{equation}
\noindent where
\begin{equation}
s(r) = - \dfrac{8\pi G}{c^2} m^2 r \left[\rho - 6 \left(\dfrac{2\xi}{m^2}\right) \nabla^2 \rho \right],
\end{equation}
\noindent is the source function.

The boundary conditions on $u(r)$ are the ones listed after Eq. \eqref{eq:trace_eq_subst_naff}. The Green function of Eq. \eqref{curvature equation, variable substitution} solves the following equation in the sense of distributions
\begin{equation}
\dfrac{d^2 G(r,r')}{dr^2} - m^2 G(r,r') = \delta (r - r'),
\end{equation}
\noindent where $\delta(r-r')$ is the Dirac delta distribution. The Green function $G(r,r')$ is used to determine the solution of the curvature equation in the form \eqref{curvature equation, variable substitution} by means of the integral $u(r) = \int^{R_S}_0 G(r,r') s(r') dr'$.
Due to the different boundary conditions, the curvature is written as a twofold solution:
\begin{equation} \label{curvature solution}
R(r) = \left\{
  \begin{array}{l l}
     R^\uparrow(r) & \quad \text{if  } r>R_S\\
     \quad \\
     R^\downarrow(r) & \quad \text{if  } 0\leq r\leq R_S   \end{array} \right. ,
\end{equation}
\noindent where $R^\downarrow(r)$ is the curvature inside the star,
\begin{equation}
R^\downarrow (r) = -\dfrac{4 \pi G}{c^2 m} \left[ \dfrac{e^{-mr}}{r} I_1(r)
+ \dfrac{2 \sinh(mr)}{r} I_2(r) \right],
\end{equation}
\noindent the quantities $I_1(r)$ and $I_2(r)$ have been computed by using the properties (\ref{conditions for fe_march})
of the mass density $\rho(r)$ and are given by
\begin{eqnarray}
I_1(r)&=& 2\left(12\xi -1\right)m^2\int^{r}_{0}\sinh(mr')r'\rho(r')dr'  -24\xi \cosh(mr)mr\rho(r),\\ \nonumber
I_2(r)&=& (12\xi-1) m^2  \int^{R_S}_r e^{-mr'}r'\rho(r')  dr' -12\xi e^{-mr} mr \rho(r)  ,
\end{eqnarray}
\noindent and $R^\uparrow(r)$ is the curvature outside the star, given by
\begin{equation} \label{curvature solution, outside star}
R^\uparrow (r) = \dfrac{2 G M_S}{c^2 r} m^2 \left( 1 - 12\xi \right) A(m,R_S) e^{-mr},
\end{equation}
\noindent with $M_S$ the mass of the spherical body and $A(m,R_S)$ a form factor defined as
\begin{equation}
A(m,R_S) = \dfrac{4\pi}{m M_S} \int^{R_S}_0 \sinh(mr) r \rho(r) dr,
\end{equation}
\noindent which will be discussed in a subsequent section.

The expression (\ref{curvature solution, outside star}) vanishes as $r\rightarrow\infty$ and it
is considered to be valid only at Solar System scales, since spacetime should assume a De Sitter metric with curvature $R_0 \neq 0$
at cosmological scales. Note also that in the limit $m \rightarrow 0$ then $R^\uparrow(r) \rightarrow 0$ for any $r > R_S$.

%%%%%%%%%%%%%%%%%%%%%%%%%%%%%%%%%%%%%%%%%%%%%%%%%%%%%%%%%%%%%%%%%%
\subsection{Solution for \texorpdfstring{$\Psi$}{Psi}}
%%%%%%%%%%%%%%%%%%%%%%%%%%%%%%%%%%%%%%%%%%%%%%%%%%%%%%%%%%%%%%%%%%

The $tt$ component of the Ricci tensor at order $\OO(1/c^2)$ is given by
\begin{equation} \label{elements, tt component}
R_{tt} = \dfrac{2\Psi'}{r}+\Psi'' + \OO\left(\dfrac{1}{c^3}\right) = \nabla^2\Psi + \OO\left(\dfrac{1}{c^3}\right).
\end{equation}
\noindent  Then, neglecting all terms smaller than $\OO(1/c^2)$ in the $tt$ component of the field equations \eqref{eq:field_equations_nmc},
and using the curvature equation \eqref{curvature equation}, the equation ruling $\Psi$ becomes:
\begin{equation} \label{psi equation}
\nabla^2\Psi = \dfrac{c^2}{3k}\rho-
\dfrac{R}{6}.
\end{equation}
This is solved outside the spherical body, $r \geq R_S$, by decomposing it into a sum,
\begin{equation} \label{psi equation, decomposition}
\nabla^2 \Psi = \nabla^2 \Psi_0 + \nabla^2 \Psi_1,
\end{equation}
\noindent where
\begin{equation} \label{psi 1 and psi 2}
\nabla^2 \Psi_0 = \dfrac{c^2}{3k} \rho~~~~,~~~~\nabla^2 \Psi_1 = -\dfrac{R}{6}.
\end{equation}
The first equation is easily solved using the divergence theorem, which gives
\begin{equation} \label{psi 0}
\Psi_0 (r) = - \dfrac{4 }{3 } \dfrac{GM_S}{c^2r}.
\end{equation}
The second one is more cumbersome: a tedious integration eventually shows that
\begin{equation} \label{psi 1}
\Psi_1 (r) = \dfrac{G M_S}{3 c^2 r} \left[ 1 - \left( 1 - 12 \xi \right) A(m,R_S) e^{-mr} \right],
\end{equation}
\noindent so that for $r \geq R_S$ the solution is
\begin{equation} \label{psi solution}
\Psi(r) = - \dfrac{G M_S}{c^2 r} \left[1 + \left(\dfrac{1}{3} - 4\xi \right)
A(m,R_S) e^{-mr}\right].
\end{equation}

%%%%%%%%%%%%%%%%%%%%%%%%%%%%%%%%%%%%%%%%%%%%%%%%%%%%%%%%%%%%%%%%%%
\subsection{Solution for \texorpdfstring{$\Phi$}{Phi}}
%%%%%%%%%%%%%%%%%%%%%%%%%%%%%%%%%%%%%%%%%%%%%%%%%%%%%%%%%%%%%%%%%%

By the same token, being the mass distribution static, the expressions
\begin{equation} \label{radial components}
R_{rr} = \dfrac{2}{r}\Phi' - \Psi'' + \OO\left(\dfrac{1}{c^3}\right)~~~~,~~~~
T_{rr} = 0,
\end{equation}
\noindent can be inserted into the $rr$ component of the field equations, along with the $f^i(R)$ expressions from \eqref{f(R) equations}, obtaining
\begin{equation} \label{phi equation}
2\Phi' - r \Psi'' - \dfrac{r R}{2} + \dfrac{2 R'}{3 m^2} -
\dfrac{4 \xi c^2}{m^2 k} \rho' = 0.
\end{equation}
\noindent This equation is easily integrated outside the spherical body, $r \geq R_S$, leading to
\begin{equation} \label{phi solution}
\Phi(r)=\dfrac{G M_S}{c^2 r} \left[ 1 - \left(\dfrac{1}{3}-4\xi\right) A(m,R_S) e^{-mr} (1 +mr) \right].
\end{equation}
\noindent In the \ac{GR} limit, $\xi = 0$ and $m \to \infty$, the exponential term in both $\Psi$ and $\Phi$ vanishes and the weak-field approximation of the Schwarzschild metric is recovered, as expected.

%%%%%%%%%%%%%%%%%%%%%%%%%%%%%%%%%%%%%%%%%%%%%%%%%%%%%%%%%%%%%%%%%%
\section{Discussion of Yukawa potential}
%%%%%%%%%%%%%%%%%%%%%%%%%%%%%%%%%%%%%%%%%%%%%%%%%%%%%%%%%%%%%%%%%%

As it has been shown, from the $tt$ component of the metric it is possible to identify a Newtonian potential plus a Yukawa perturbation:
\begin{equation} \label{yukawa potential}
U(r) = - \dfrac{G M_S}{r} \left( 1 + \alpha A(m,R_S) e^{-r/\lambda} \right),
\end{equation}
\noindent defining the characteristic length $\lambda = 1/m$ and the strength of the Yukawa addition
\begin{equation} \label{alpha as the yukawa potential strength}
\alpha = \dfrac{1}{3} - 4\xi ,
\end{equation}
\noindent so that, if $\xi = 0$, the Yukawa strength for pure $f(R)$ theories from chapter \ref{cap:fofR} is obtained, $\alpha = 1/3$; it should also be noticed that a positive \ac{NMC} (as assumed in Refs. \cite{stelobserv,gravcollapse}) yields $\alpha \leq 1/3$. Strikingly, a NMC with $\xi = 1/12$ cancels the Yukawa contribution.

%%%%%%%%%%%%%%%%%%%%%%%%%%%%%%%%%%%%%%%%%%%%%%%%%%%%%%%%%%%%%%%%%%
\subsection{The form factor \texorpdfstring{$A(m,R_S)$}{A(m,Rs)}}
%%%%%%%%%%%%%%%%%%%%%%%%%%%%%%%%%%%%%%%%%%%%%%%%%%%%%%%%%%%%%%%%%%

As defined before, the form factor is
\begin{equation} \label{form factor A(m,R_S)}
A(m,R_S) = \dfrac{4\pi}{m M_S} \int_0^{R_S} \sinh(mr) r\rho(r)dr.
\end{equation}
This dimensionless form factor was found by integrating the field equations of NMC gravity but it is not specific of the NMC gravity nor of $f(R)$ theories of this kind, as it was shown in section \ref{sec:isl}. This form factor can then be evaluated in several ways, according to the function of mass density $\rho(r)$. Taking the limit of a point source, $r \to 0$ allows to expand around $m r \ll 1$, so that $\sinh \left( m r \right) \approx m r [1 + (mr)^2/6]$ and
\begin{equation}
A(m,R_S) \approx \dfrac{4\pi}{M_S} \int^{R_S}_0 \left[1 + \dfrac{(mr)^2 }{6}\right] r^2 \rho(r) dr
=  1 + \dfrac{2m^2\pi}{ 3M_S} \int^{R_S}_0 r^4 \rho(r) dr \sim 1.
\end{equation}
\noindent This can be verified explicitly by making all computations with a test mass density (such as a uniform profile) and, in the end, taking the limit $R_S \to 0$. Indeed, taking
\begin{equation}
\rho_0 = \dfrac{3M_S}{4\pi R_S^3}
\end{equation}
\noindent then
\begin{equation} \label{form factor A(m,R_S) constant}
A(m,R_S) =3 \dfrac{mR_S \cosh (mR_S) - \sinh (mR_S)}{(m R_S)^3} ,
\end{equation}
\noindent which admits the limiting cases
\begin{eqnarray} \label{form factor A(m,R_S) constant limiting}
A(m,R_S) &\approx & 1 + \dfrac{(mR_S)^2 }{ 10} \sim 1 ~~,~~mR_S \ll 1, \\ \nonumber A(m,R_S) &\approx & \dfrac{3}{2}\dfrac{e^{mR_S}}{(mR_S)^2} ~~,~~mR_S \gg 1.
 \end{eqnarray}
If the central body is the Sun (with radius $R_\odot$), the more accurate density NASA profile \cite{NASAprofile} can instead be considered (which obeys condition $\rho(R_\odot)=0$, while $(d\rho\slash dr)(R_\odot) \simeq 0$),
\begin{equation}\label{NASA density profile}
\rho(r) = \rho_0 \bigg[  1 - 5.74 \left(\dfrac{r}{R_\odot}\right)+ 11.9\left(\dfrac{r}{R_\odot}\right)^2 - 10.5\left(\dfrac{r}{R_\odot}\right)^3 + 3.34\left(\dfrac{r}{R_\odot}\right)^4\bigg] ,
 \end{equation}
\noindent obtaining
\begin{eqnarray} \label{form factor A(m,R_S) NASA}
A(m,R_\odot) &=& x^{-7} [4.6 \times 10^4 x + 2.1 \times 10^3 x^3 +  (2.7 \times 10^4 + 131 x^2) x \cosh x - \\ \nonumber &&  (7.3 \times 10^4 + 3.6 \times 10^3x^2 - 14.6 x^4) \sinh x ],
\end{eqnarray}
\noindent (with $x= mR_\odot$, for brevity), with the limiting cases
\begin{eqnarray} \label{form factor A(m,R_S) NASA limiting}
A(m,R_\odot) &\approx & 1 + 6 \times 10^{-2} (m R_\odot)^2 \sim 1 ~~,~~mR_\odot \ll 1, \nonumber \\ A(m,R_\odot) &\approx & \ 7.3 \dfrac{e^{mR_\odot}}{(mR_\odot)^3 }~~,~~mR_\odot \gg 1.
 \end{eqnarray}

\begin{figure}[ht]
\centering
\includegraphics[width=\textwidth]{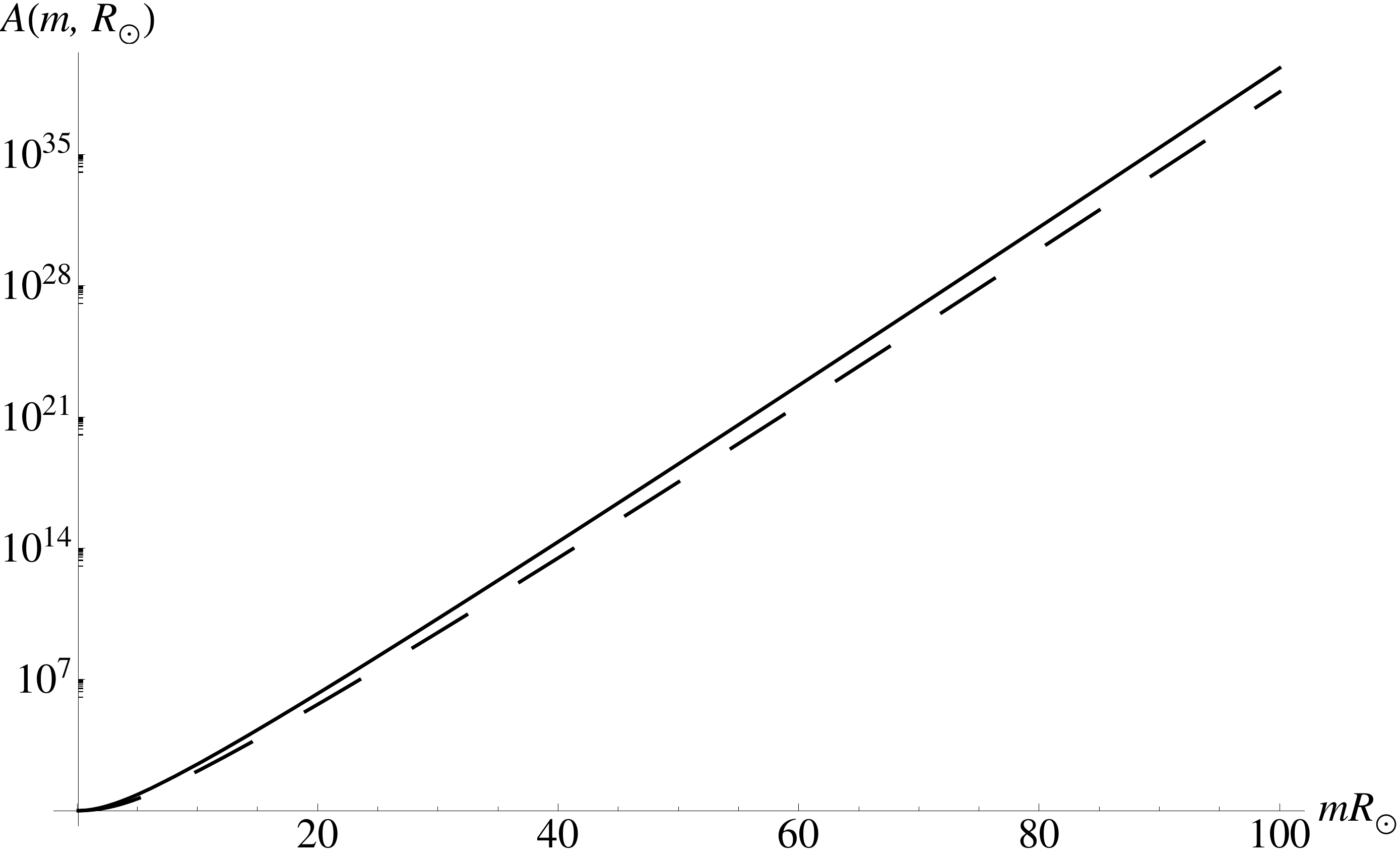}
\caption{Numerical plot of form factor $A(m,R_\odot)$ for a constant density (full) and fourth-order density profile for the Sun, Eq. (\ref{NASA density profile}) (dashed).}
\label{fig:formfactors}
\end{figure}

Both forms for $A(m,R_\odot)$ are plotted in Fig. \ref{fig:formfactors}, showing that it grows with $mR_\odot$. Although this effectively boosts the form factor for values of lengthscale $\lambda \ll R_\odot$, the contribution from the Yukawa term in Eq. (\ref{yukawa potential}) is nevertheless suppressed by the factor $\exp(-r/\lambda)$.

%%%%%%%%%%%%%%%%%%%%%%%%%%%%%%%%%%%%%%%%%%%%%%%%%%%%%%%%%%%%%%%%%%
\subsection{\ac{PPN} Parameters}
%%%%%%%%%%%%%%%%%%%%%%%%%%%%%%%%%%%%%%%%%%%%%%%%%%%%%%%%%%%%%%%%%%

Similarly to the present considerations, the \ac{PPN} formalism posits an expansion of the metric elements and other quantities (energy-momentum tensor, equations of motion, {\it etc}.) in powers of $1/c^2$, as it was introduced in section \ref{sec:ppn}.

Clearly, such a formalism is incompatible with the presence of a Yukawa term in the gravitational potential, since the latter cannot be expanded in powers of $1/r$; furthermore, the discussion after Eq. (\ref{curvature solution, outside star}) highlights that, in the limit $m\to 0$, the background cosmological curvature must be considered and the asymptotically flat {\it Ansatz} (\ref{metric}) for the metric cannot be assumed: this was performed in section \ref{sec:impact_nmc}, as already mentioned.

Nevertheless, for consistency, what happens if the condition $mr \ll 1$ is valid throughout the region of interest ({\it e.g.} the Solar System) can be considered: in this case, the metric (\ref{metric}) with the solutions (\ref{psi solution}) and (\ref{phi solution}) is well approximated by
\begin{equation} \label{metriclight}
ds^2 = -\left[1 - \dfrac{2G M_S}{c^2 r} \left(\dfrac{4}{3} - 4\xi \right) \right]c^2dt^2+ \left[1+\dfrac{2G M_S}{c^2 r} \left( \dfrac{2}{3}+ 4\xi \right)\right] dr^2 + r^2 d\Omega^2,
\end{equation}
\noindent which yields
\begin{equation}\label{gamma}
\gamma= \dfrac{1 }{ 2} \dfrac{1 + 6\xi }{ 1 - 3\xi}.
\end{equation}
In the absence of a \ac{NMC}, $\xi = 0 $, this yields $\gamma = 1/2$, a strong departure from \ac{GR} that is disallowed by current experimental bounds, $\gamma = 1 + (2.1 \pm 2.3) \times 10^{-5}$ \cite{cassini}. This apparent disagreement between $f(R)$ theories and observations was already noted (as discussed in section \ref{sec:solar_sys_impact}), and can be avoided if the additional degree of freedom arising from a non-linear $f(R)$ function is massive enough.

The expression above appears to show that a \ac{NMC} allows $f(R)$ theories to remain compatible with observations, as long as $\xi=1/12$ --- which is just a restatement of the previously obtained result. Again, the path towards obtaining the $\gamma$ \ac{PPN} parameter depicted above is presented for illustration only, as it relies on an approximation of a Yukawa perturbation and disregards the fact that, in the limit $mr \ll 1$, the background cosmological dynamics cannot be neglected. As such, no conclusions can be drawn from comparison with the experimental bound on $\gamma$ mentioned above.

%%%%%%%%%%%%%%%%%%%%%%%%%%%%%%%%%%%%%%%%%%%%%%%%%%%%%%%%%%%%%%%%%%
\section{Experimental constraints to NMC gravity parameters}
%%%%%%%%%%%%%%%%%%%%%%%%%%%%%%%%%%%%%%%%%%%%%%%%%%%%%%%%%%%%%%%%%%

\begin{figure}[ht]
\centering
\includegraphics[width=\textwidth]{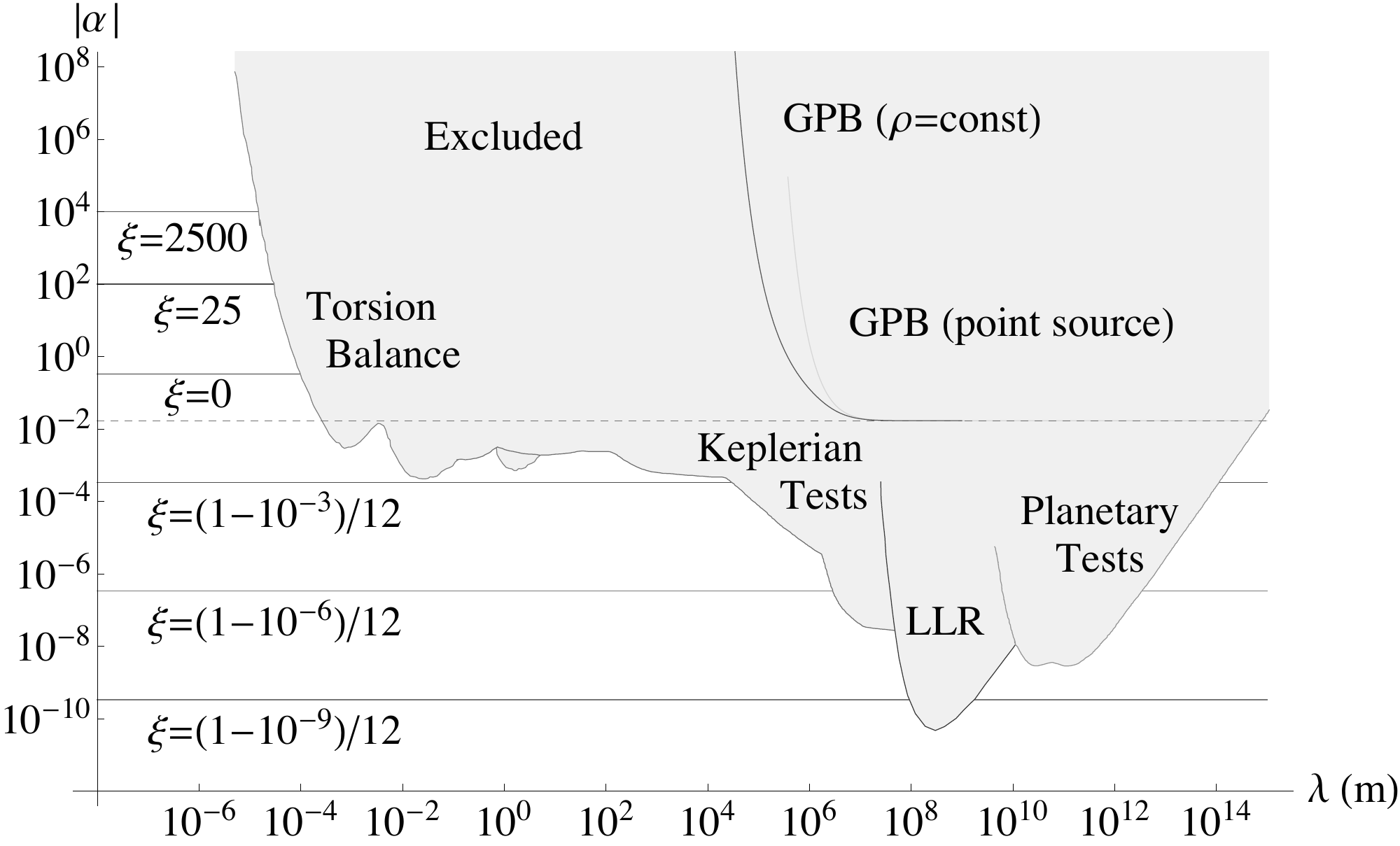}
\caption{Yukawa exclusion plot for $\alpha$ and $\lambda$. Adapted from Refs. \cite{adelberg1,salumbides}.}
\label{fig:exclusionplot}
\end{figure}

The Yukawa potential \eqref{yukawa potential} is not new in physics as an alternative way to account for deviations from Newtonian gravity or other forces of nature, as it was shown in section \ref{sec:isl}. Fig. \ref{fig:exclusionplot} shows the exclusion plot for the phase space $(\lambda,\alpha)$, which may be used to constrain the phase space of the model (\ref{eq:action_nmc}) under scrutiny.

In doing so, it is important to recall that the results obtained in Eqs. (\ref{psi solution},\ref{phi solution},\ref{yukawa potential}) are not exact, but only accurate to order $\OO(c^{-2})$, and are based upon the assumption of a perturbation to a Minkowski metric: a future analysis should expand this framework to also include terms $\OO(c^{-4})$, as well as establish some matching criteria between the static, spherically symmetric spacetime here considered and the evolving background spacetime \cite{matching} (see also Ref. \cite{darkmatter}). Indeed, Ref. \cite{Clifton} has found that in $f(R)$ theories, $\OO(c^{-4})$ terms can arise that are not exponentially suppressed, and as such may play a role at large distances, particularly if the $\OO(c^{-2})$ Yukawa interaction here obtained is short-ranged.

From Fig. \ref{fig:exclusionplot} (as discussed after \eqref{alpha as the yukawa potential strength}), it is seen that a \ac{NMC} with $\xi = 1/12$ cancels this contribution, as shown by the values of $|\alpha| \to 0$ overlaid on the exclusion plot. Also, it should be noticed that large values of $\xi$ lead to a large, negative strength $\alpha \sim -4\xi$ (the cases $\xi =25$ and $\xi = 2500$ are shown).

\begin{figure}[ht]
\centering
\includegraphics[width=\textwidth]{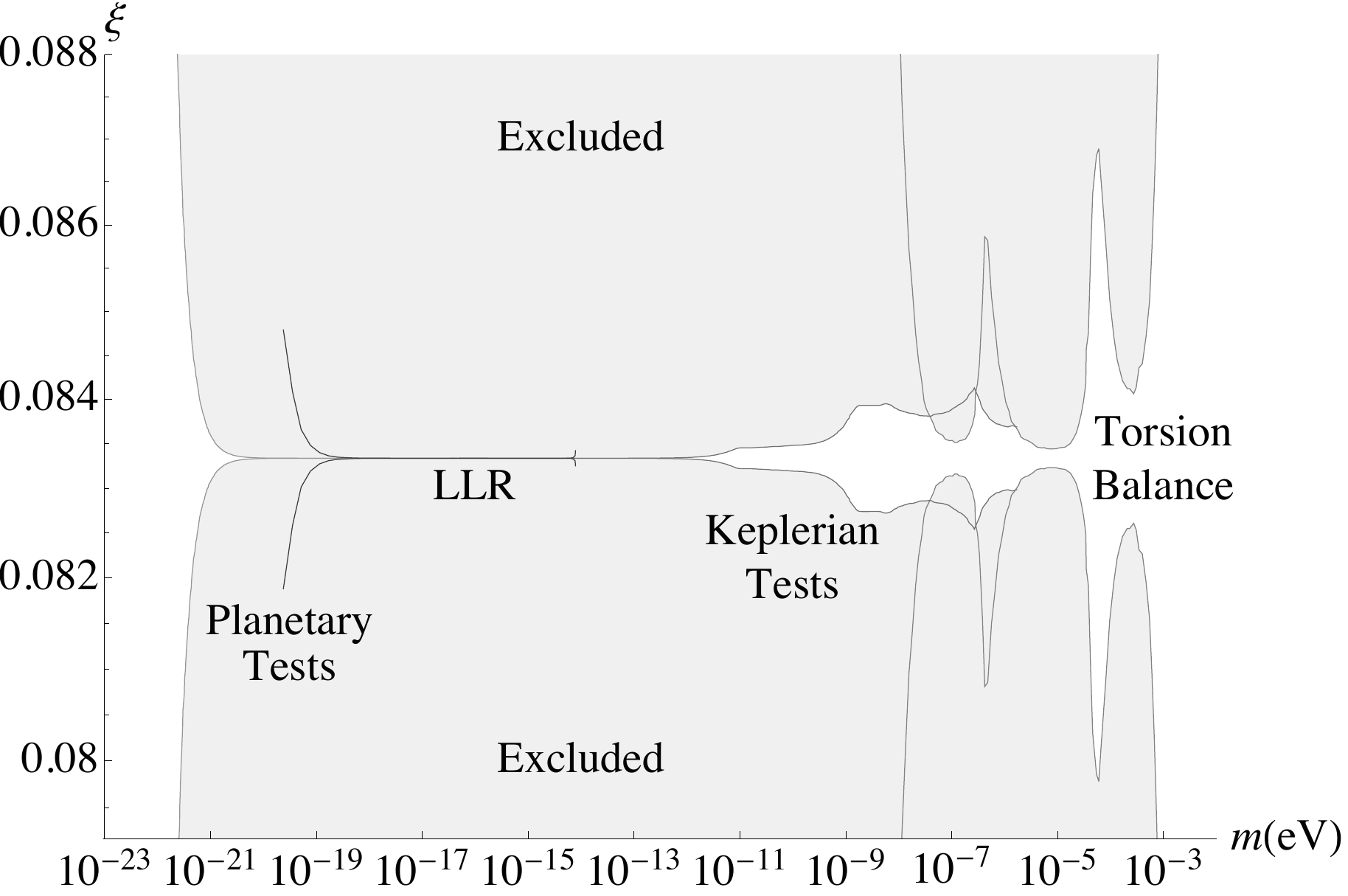}
\caption{Exclusion plot for the dimensionless relative strength $\xi$ and characteristic mass scale $m$.}
\label{fig:exclusionplotnmccoord}
\end{figure}

This plot may be transformed into the phase space $(m,\xi)$ of the model under scrutiny, \eqref{f(R) equations}, using
\begin{equation} \label{transformations, nmc graphic}
m = \frac{1}{\lambda}, \qquad \qquad
\xi = \frac{1}{12}-\frac{\alpha}{4},
\end{equation}
\noindent to get the exclusion plot depicted in Fig. \ref{fig:exclusionplotnmccoord}.

\begin{figure}[ht]
\centering
\includegraphics[width=\textwidth]{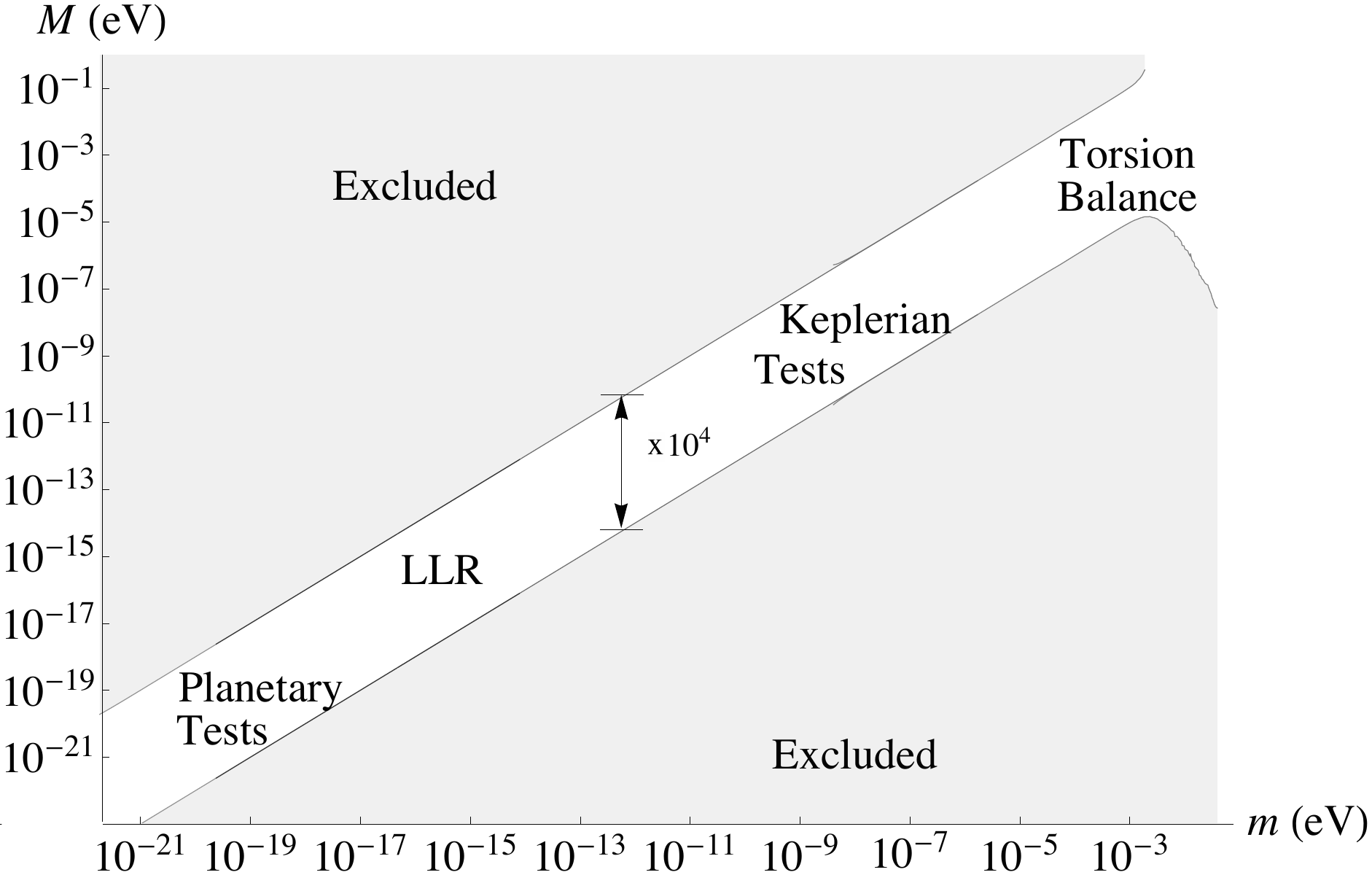}
\caption{Exclusion plot for the characteristic mass scales $M$ and $m$.}
\label{fig:mexclusionplotnmccoord}
\end{figure}

Further insight is obtained by casting the \ac{NMC} presented in Eq. \eqref{f(R) equations} as
\begin{equation}\label{transformation, nmc parameter}
f^2(R) = \frac{R}{6 M^2},
\end{equation}
\noindent so that it is characterized by a distinct mass scale $M$, instead of the relative strength parameter $\xi$: by making the transformation $\xi= (m/M)^2/12$, the suggestive form
\begin{equation} \label{transformation, new alpha for new nmc parameter}
\alpha = \dfrac{1}{3} \left[ 1 - \left( \dfrac{m}{M} \right)^2 \right],
\end{equation}
\noindent is thus obtained, which, inverting, allows to plot the exclusion plot for the phase space $(m,M)$ in Fig. \ref{fig:mexclusionplotnmccoord}.

%%%%%%%%%%%%%%%%%%%%%%%%%%%%%%%%%%
%%%%%%%%% BEGIN PICTURES %%%%%%%%%
%%%%%%%%%%%%%%%%%%%%%%%%%%%%%%%%%%

%%%%%%%%%%%%%%%%%%%%%%%%%%%%%%%%%%
%%%%%%%%%% END PICTURES %%%%%%%%%%
%%%%%%%%%%%%%%%%%%%%%%%%%%%%%%%%%%

Figs. \ref{fig:exclusionplot}-\ref{fig:mexclusionplotnmccoord} show us that, if $m$ falls within the range $10^{-22} ~{\rm eV} < m < 1~{\rm meV}$ (corresponding to lengthscales $\lambda$ ranging from the millimeter to Solar System scales), then the strong constraints available on the Yukawa strength, $|\alpha| \ll 1$, require that $\xi \sim 1/12$ --- or, equivalently, that both characteristic mass scales are very similar, $m \sim M$.

%%%%%%%%%%%%%%%%%%%%%%%%%%%%%%%%%%%%%%%%%%%%%%%%%%%%%%%%%%%%%%%%%%
\subsection{Geodetic precession}
%%%%%%%%%%%%%%%%%%%%%%%%%%%%%%%%%%%%%%%%%%%%%%%%%%%%%%%%%%%%%%%%%%

In this section it is assumed that the Earth can be approximated as a spherically symmetric body.
In order to assess the impact of the obtained expression for $\Phi(r)$, Eq. \eqref{phi solution}, we now consider a gyroscope in a circular orbit with radius $r$ around the Earth.
According to section \ref{sec:geodetic}, the intrinsic angular momentum vector $S^\mu = \left(S^0, \mathbf{S} \right)$ precesses according to the equation of parallel transport:
\begin{equation}
\frac{dS^\mu}{d\tau} = -\Gamma^\mu_{\nu\sigma}S^\nu\frac{dx^\sigma}{d\tau},
\end{equation}
\noindent where $\tau$ is the proper time. For convenience, the metric tensor around the Earth is written in rectangular isotropic coordinates,
\begin{equation} \label{metric in rectangular isotropic coordinates}
ds^2 = -\left[ 1 - \frac{2GM_S}{c^2r}\left( 1 + \alpha A(m,R_S)e^{-mr} \right)\right]c^2dt^2 + 
 \left[ 1 + \frac{2GM_S}{c^2r}\left( 1 -  \alpha A(m,R_S) e^{-mr} \right)\right] dV^2.
\end{equation}
The standard method of computation of gyroscope precession in \ac{GR} yields for the secular part of
$d\mathbf{S}\slash dt$ in \ac{NMC} gravity, in the slow motion and weak field approximation,
\begin{equation}
\left( \dfrac{d\mathbf{S}}{dt} \right)_{{\rm sec}} = \frac{3}{2}\frac{G M_S}{c^2r^3} \left[ 1 - \frac{\alpha A(m,R_S)}{3}(1 + mr)e^{-mr} \right]\left( \mathbf{r} \times \mathbf{v} \right) \times \mathbf{S},
\end{equation}
\noindent where $\mathbf{r}$ is the radius vector of the center of mass of the gyroscope and $\mathbf{v}$ is its velocity vector.
Imposing the equality between the acceleration $v^2\slash r$ of the center of mass of the gyroscope and the sum of the Newton plus Yukawa forces per unit mass yields
\begin{equation}
vr = \sqrt{G M_S r [ 1 + \alpha A(m,R_S)(1 + mr)e^{-mr} ] }.
\end{equation}
\noindent Since $\left(d\mathbf{S} \slash dt\right)_{{\rm sec}} = \boldsymbol{\Omega}_G \times \mathbf{S}$, the angular velocity vector $\boldsymbol{\Omega}_G$ of geodetic precession is given by
\begin{equation}
\boldsymbol{\Omega}_G = \frac{3}{2}\frac{\left(G M_S \right)^{3\slash 2}}{c^2r^{5\slash 2}} \left[ 1 + \alpha A(m,R_S)(1+mr)e^{-mr} \right]^{1\slash 2} \times \left[ 1 - \frac{\alpha A(m,R_S)}{3}(1+mr)e^{-mr} \right] \mathbf{n},
\end{equation}
\noindent where $\mathbf{n}$ is the unit vector perpendicular to the plane of the orbit. If $\xi=0$, the above expression reduces to the case for $f(R)$ models, as expected \cite{naff}.

%%%%%%%%%%%%%%%%%%%%%%%%%%%%%%%%%%%%%%%%%%%%%%%%%%%%%%%%%%%%%%%%%%
\subsubsection{Gravity Probe B results}
%%%%%%%%%%%%%%%%%%%%%%%%%%%%%%%%%%%%%%%%%%%%%%%%%%%%%%%%%%%%%%%%%%

As it was already mentioned, the final results of the \ac{GPB} experiment report an accuracy of 0.28\% in the measurement of geodetic precession \cite{GPB}. This corresponds to the following constraint on \ac{NMC} gravity parameters:
\begin{equation} \label{GPB constrain, general}
\left\vert \frac{\Omega_G - \Omega_G^{{\rm GR}}}{\Omega_G^{{\rm GR}}} \right\vert < 0.0028,
\end{equation}
\noindent where only the modulus of angular velocity is considered, and $\Omega_G^{{\rm GR}}$ denotes the value of
geodetic precession in \ac{GR}. Substituting the expression of \ac{NMC} geodetic precession in this constraint we find
\begin{equation} \label{GPB constrain, NMC}
\Big\vert \sqrt{ 1 + \alpha A(m,R_S) (1+mr)e^{-mr} } \left[ 1 - \frac{\alpha A(m,R_S)}{3}(1+mr)e^{-mr} \right] - 1 \Big\vert  < 0.0028.
\end{equation}
Defining $x \equiv \alpha A(m,R_S) (1 + mr) e^{-mr}$, this is written as
\begin{equation}
\Big\vert \sqrt{ 1 +x } \left( 1 - \frac{x}{3} \right) - 1 \Big\vert  < 0.0028.
\end{equation}
If $x \gg 1$, then $|x| < 0.04$, which is contradictory. Since substitution shows that $x \sim 1$ breaks the above relation, the only natural constraint left is $x \ll 1$, so that a first order expansion of the above yields $|x| < 0.0168$. This last condition is translated as
\begin{equation} \label{GPB, alpha for small x}
|\alpha | < \dfrac{0.0168}{1 + mr}\dfrac{e^{mr}}{A(m,R_S)}.
\end{equation}
In order to satisfy the assumption (\ref{conditions for fe_march}) of continuity of mass density and its derivative across the surface of the Earth, it is possible to model the density with a constant value in an interior region ({\it i.e.} mantle plus core) and a sharp transition in a thin crustal layer. When the thickness of the latter tends to zero, the form factor $A(m,R_S) $ converges to the value corresponding to the uniform density model, Eq. (\ref{form factor A(m,R_S) constant limiting}), hence the inequality (\ref{GPB, alpha for small x}) reads
\begin{eqnarray} \label{GPB, alpha for small x limiting}
|\alpha | &<& 0.0168~~,~~mR_\oplus \ll 1 ~~, \\ \nonumber
|\alpha | &<& 0.0112\dfrac{mR_\oplus^2}{ r}e^{m(r-R_\oplus)}~~,~~mR_\oplus \gg 1 ~~,
\end{eqnarray}
\noindent where $R_\oplus \approx 6371$ km is the radius of the Earth. Knowing that the \ac{GPB} satellite orbits the Earth at a height of $\sim 650$ km, this condition can be plotted in the $(\lambda,\alpha)$ exclusion plot, as shown in Fig. \ref{fig:exclusionplot}. It is found that the condition is well-within the already excluded phase space, so that the current bounds on geodetic precession do not add any new constraint on the model parameters.

%%%%%%%%%%%%%%%%%%%%%%%%%%%%%%%%%%%%%%%%%%%%%%%%%%%%%%%%%%%%%%%%%%
\subsubsection{Measurement of the LAGEOS II perigee precession}
%%%%%%%%%%%%%%%%%%%%%%%%%%%%%%%%%%%%%%%%%%%%%%%%%%%%%%%%%%%%%%%%%%

A recent analysis of the perigee precession of the LAGEOS II satellite reported a much stronger constraint on the strength of the Yukawa perturbation, $\vert\alpha \vert\simeq \vert(1.0\pm 8.9)\vert\times 10^{-12}$, at a range $\lambda = 1\slash m = 6081$ km, very close to one Earth radius \cite{lageos}: a striking improvement over previous Earth-LAGEOS and Lunar-LAGEOS measurements (at the level of $10^{-5}$ and $10^{-8}$), and comparable to the Lunar Laser Ranging constraint on $\alpha$ for $\lambda \sim 60 R_\oplus$ \cite{LLR}.

Non-gravitational perturbations, mainly
thermal perturbative effects, can strongly affect the precession of the perigee of LAGEOS II: in \cite{lageos}, solar radiation pressure
and Earth's albedo are taken into account, while Rubincam and Yarkovsky-Schach (YS) thermal effects (which need the satellite spin modeling)
have not been considered. Nevertheless, the residuals in the perigee rate of the satellite are fitted with a linear trend (which represents the secular
total \ac{GR} precession) plus four periodic terms which correspond to the main spectral lines of the unmodeled YS effect \cite{lageos}.

This said, if the impressive bound on $\alpha$ quoted above is indeed confirmed, no qualitative changes occur in the previous analysis: as long as the Yukawa coupling strength lies below unity sufficiently, then the \ac{NMC} parameter must be $\xi \sim 1/12 \rightarrow M \sim m$, so that lowering the upper bound on the former only brings the two mass scales of the functions $f^1(R)$ and $f^2(R)$ closer together.

\cleardoublepage

% %%%%%%%%%%%%%%%%%%%%%%%%%%%%%%%%%%%%%%%%%%%%%%%%%%%%%%%%%%%%%%%%%%%%%%
% The Introduction:
% %%%%%%%%%%%%%%%%%%%%%%%%%%%%%%%%%%%%%%%%%%%%%%%%%%%%%%%%%%%%%%%%%%%%%%
\chapter{Discussion and Outlook}
\label{cap:conclusions}

In this work it has been computed the effect of a \ac{NMC} model, specified by \eqref{f(R) equations}, in a perturbed weak-field Schwarzschild metric, as depicted in Eq. (\ref{psi solution}) and (\ref{phi solution}). In the weak-field limit, this translates into a Yukawa perturbation to the usual Newtonian potential, with characteristic range and coupling strength
\begin{equation}
\lambda = \dfrac{1}{m}~~~~,~~~~\alpha = \left( \dfrac{1}{3} - 4\xi \right) = \dfrac{1}{3} \left[ 1 - \left( \dfrac{m}{M} \right)^2 \right]~~.
\end{equation}
\noindent This result is quite natural and can be interpreted straightforwardly: a minimally coupled $f(R)$ theory introduces a new massive degree of freedom (as hinted by the equivalence with a scalar-tensor theory \cite{analogy1,analogy2,analogy3}), leading to a Yukawa contribution with characteristic lengthscale $\lambda = 1/m$ and coupling strength $\alpha = 1/3$.

The introduction of a \ac{NMC} has no dynamical effect in the vacuum, as there is no matter to couple the scalar curvature to: as a result, it is not expected any modification in the range of this Yukawa addition. Conversely, a \ac{NMC} has an impact on the description of the interior of the central body leading to a correction to the latter's coupling strength (which has a negative sign since $\LL_m= -\rho$), as mentioned in the beginning of chapter \ref{cap:nmc}.

Using the available experimental constraints, it was found that, for $10^{-22}~{\rm eV} < m < 1~{\rm meV}$ ({\it i.e.} the range $ 10^{-4}~{\rm m} < \lambda < 10^{16}~{\rm m}$), where $|\alpha| \ll 1$, then the \ac{NMC} parameter must be $\xi \sim 1/12$ or, equivalently, both mass scales $m$ and $M$ of the non-trivial functions $f^1(R)$ and $f^2(R)$ must be extremely close.

If this is the case, the latter relation is not interpreted as an undesirable fine-tuning, but instead is suggestive of a common origin for both non-trivial functions $f^1(R)$ and $f^2(R)$, in line with the argument stating that the model \eqref{eq:action_nmc} should arise as a low energy phenomenological approximation to a yet unknown fundamental theory of gravity.

Furthermore, one notices that the result $\xi = 1/12$ is directly equivalent to the action
\begin{equation}\label{model2}
S=\int  \left(1+\dfrac{R}{6m^2}\right) \left(\kappa R + \LL_m\right)
\sqrt{-g} d^4x,
\end{equation}
\noindent and the main result of this work can be recast in a more appealing way: if an analytical (around $R=0$) NMC model leads to a massive degree of freedom with a short range which falls within the currently observable region $1~{\rm mm} \lesssim \lambda \lesssim 10^{15} ~{\rm m}$, then it must correspond to a global factorisation of the Einstein-Hilbert Lagrangian, as shown above. Further studies based on this insight, and its extension to higher order couplings, can be realised by considering the so-called $F(R,\LL_m)$ set of theories \cite{FRL}.

Conversely, for values of $m$ (or $\lambda$) away from the range mentioned above the Yukawa coupling strength, $\alpha$ can be much larger than unity, so that $\xi $ can assume any value and the mass scales $m$ and $M$ can differ considerably.

In particular, the Starobinsky inflationary model, which requires the much heavier mass scale $m \approx 3 \times 10^{13}$ GeV $\sim 10^{-6} M_P$, manifests itself at a lengthscale $\lambda \sim 10^{-29}~{\rm m}$. This implies that the generalized preheating scenario posited in section \ref{sec:reheating} and in Ref. \cite{reheating}, which requires $1<\xi<10^4$, is thus completely allowed by experiment and unconstrained by this work.

By computing the perturbation induced on geodetic precession, it was found that no significant new constraint arises, as this is already included in the existing Yukawa exclusion plot. Furthermore, even considering the much improved precision claimed in a recent study of LAGEOS II --- or, for that matter, any further refinement of $|\alpha| \ll 1$ $-$, no qualitatively new results arise, since this only bridges the gap between $m $ and $M$ ({\it i.e.} narrows the value of $|\xi - 1/12|$).

%\begin{comment}
Finally, a word is due for the so-called chameleon mechanism, first posited in Ref. \cite{chameleon1,chameleon2,chameleon3,chameleon4}, as discussed in Ref. \cite{naff} in relation to $f(R)$ theories. This non-linear effect goes beyond the linear expansion of the modified field equations, and relies on the equivalence between $f(R)$ theories and a scalar-tensor theory with a scalar field $\phi$ proportional to $f_R$, which appears non-minimally coupled to the matter Lagrangian density (in the Einstein frame) \cite{analogy1,analogy2,analogy3,felice}.

As it turns out, the effective potential of this scalar field can be written as $V_{eff}(\phi) = V(\phi) + e^{a\phi}\rho$ (with $a$ an appropriate constant), so that the position of its minimum depends on the density $\rho$, and the mass for the scalar field grows with the density: this is particularly relevant in a cosmological context, where the low background density yields a light, long-ranged field.
%\end{comment}

Given the above, Ref. \cite{naff} speculates that further computations allowing for this non-linear effect could lead to different constraints on the mass scale $m$ of the adopted quadratic form for $f^1(R)$: quite naturally, the inclusion of a direct coupling between curvature ({\it vis-\`a-vis} the scalar field) and matter only heightens this possibility.

Clearly, this prompts for a future study of the relation between this chameleon mechanism and a \ac{NMC} model, in the framework of its equivalence with a multi-scalar-tensor theory as shown in section \ref{sec:scalar-fields}. Nonetheless, one should notice that for $\xi=1/12$ the chameleon mechanism is not necessary.

\cleardoublepage

%\addcontentsline{toc}{chapter}{Bibliography}
\bibliographystyle{apsrev4-1,custom}
%\bibliography{02.biblio}
\bibliography{tese}
\addcontentsline{toc}{chapter}{Bibliography}
%\input{100.Bibliography/bibliography.tex}

%\cleardoublepage

\begin{comment}
\begin{appendices}
	\begin{appendix}
		\pagenumbering{bychapter}
		\input{Appendices/appendixA.tex}   
		\cleardoublepage
	\end{appendix}
\end{appendices}
\end{comment}

\end{document}